\documentclass[preprints,article,accept,moreauthors,pdftex]{Definitions/mdpi}

\usepackage{amsmath,amssymb,braket, bm}
\usepackage{ulem}
\newcommand{\ham}{\mathcal{\hat{H}}_{\text{LL}}}
\firstpage{1} 
\pubvolume{xx}
\issuenum{1}
\articlenumber{5}
\pubyear{2020}
\copyrightyear{2020}
\history{Received: date; Accepted: date; Published: date}

\pdfoutput=1

\Title{
Quantum dark solitons in the 1D Bose gas: From single to double dark-solitons}

\Author{Kayo Kinjo$^{1}$, Eriko Kaminishi$^{2}$, Takashi Mori$^{3}$, Jun Sato$^{4}$, Rina Kanamoto$^{5}$ and Tetsuo Deguchi$^{6}$
}

\AuthorNames{Kayo Kinjo, Eriko Kaminishi, Takashi Mori, Jun Sato, Rina Kanamoto and Tetsuo Deguchi}

\address{%
$^{1}$ \quad Department of Physics, Graduate School of Humanities and Sciences, Ochanomizu University; 
kayo.kinjo1@gmail.com\\
$^{2}$ \quad Quantum Computing Center, Keio University; 
kaminishi@keio.jp\\
$^{3}$ \quad RIKEN Center for Emergent Matter Science (CEMS);   takashi.mori.fh@riken.jp \\
$^{4}$ \quad Faculty of Engineering, Tokyo Polytechnic University;   j.sato@eng.t-kougei.ac.jp\\
$^{5}$ \quad Department of Physics, Meiji University; kanamoto@meiji.ac.jp\\
$^{6}$ \quad Department of Physics, Faculty of Core Research, Ochanomizu University; 
deguchi@phys.ocha.ac.jp
}

\corres{Correspondence: deguchi@phys.ocha.ac.jp
}

\abstract{
We study quantum double dark-solitons by constructing corresponding quantum states in the Lieb-Liniger model for the one-dimensional Bose gas.
Here we expect that the Gross-Pitaevskii (GP) equation should play a central role in the long distance mean-field behavior of the 1D Bose gas. We first introduce novel quantum states of a single dark soliton with a nonzero winding number. We show them by exactly evaluating not only the density profile but also the profiles of the square amplitude and phase of the matrix element of the field operator between the $N$-particle and $(N-1)$-particle states. 
For elliptic double dark-solitons, the density and phase profiles of the corresponding states almost perfectly agree with those of the classical solutions, respectively, in the weak coupling regime.
We then show that the scheme of the mean-field product state is quite  effective for the quantum states of double dark solitons. Assigning the ideal Gaussian weights to a sum of the excited states with two particle-hole excitations we obtain double dark-solitons of distinct narrow notches with different depths. We suggest that the mean-field product state should be well approximated by the ideal Gaussian weighted sum of the low excited states with a pair of particle-hole excitations. The results of double dark-solitons should be fundamental and useful for constructing quantum multiple dark-solitons.}


\begin{document}


\section{Introduction}

The Gross-Pitaevskii (GP) equation plays a  fundamental role in the study of  interacting nonuniform gases realized in cold atoms such as the Bose Einstein condensation (BEC)  achieved in traps\cite{ptaevskii2003BEC}. 
It is expressed in terms of the classical complex scalar field $\psi$  as follows.  
\begin{align}
i \frac {\partial } {\partial t} \psi =- \frac {\partial ^2} {\partial x^2 } \psi + 2c |\psi|^2\psi -\mu\psi . \label{eq:GPE}  
\end{align}
The non-linearity of the GP equation gives rise to a novel series of physical phenomena such as grey solitons. The GP equation has an analogous form as the Scr{\"o}dinger equation with a nonlinear term of the fourth degree. In one dimension it corresponds to the so-called nonlinear Scr{\"o}dinger (NS) equation, which is one of the fundamental integrable nonlinear equations characterized by multi-soliton solutions\cite{faddeev2007hamiltonian}. If the coupling constant is negative the NS equation has localized soliton solutions, called bright solitons, while if it is positive the NS equation has dark or grey solitons exhibiting notches in the constant density solution \cite{tsuzuki}. Here we remark that bright solitons are  actively studied in association with optical fibers\cite{abdullaev1993optical},
while dark solitons are observed 
in cold atomic experiments \cite{PhysRevLett.83.5198, PhysRevLett.101.130401, becker2008oscillations}. 

It should be quite an interesting but nontrivial problem to study how far the GP equation can describe the behavior of quantum many-body systems such as the BECs. The GP equation is a partial differential equation for the wave function as a   complex-valued order parameter, which is simply a classical complex scalar field. It does not contain any quantum fluctuations in a  straightforward manner.  
In order to study the relevance or applicability of the GP equation explicitly we may consider the Lieb-Liniger model. It describes the Bose particles in one dimension interacting through the delta-function interaction potentials. We briefly call it the one-dimensional (1D) Bose gas. The GP equation corresponds to the Heisenberg equation of motion of the field operator for the second-quantized Hamiltonian of the Lieb-Liniger model where we assume that the system size is finite.

It is known that the NS equation plays an important role in the quantum inverse scattering method, which can be regarded as a quantum version of the inverse scattering method for classical integrable systems. 
However, it is not trivial to construct such quantum states that can be related to solitons in the NS equation even through a certain kind of limiting procedure. 
 In the attractive case, bright solitons are realized in terms of string-type bound solutions of the Lieb-Liniger model \cite{WadachiSakagami}. However, in the repulsive case, the development of the studies was slightly more involved.  
 Ishikawa and Takayama argued that the dispersion relation of the type II excitation of the 1D Bose gas in the repulsive case corresponds to that of a dark (or grey)  soliton solution\cite{doi:10.1143/JPSJ.49.1242(Ishikawa-Takayama)}. 
Here we remark that some aspects of the connection were also addressed by several authors previously. However, any quantum state which can be explicitly related to a dark or grey soliton solution of the GP equation was not known for some period of time. Recently, two different approaches to the quantum states associated with dark soliton solutions have appeared. First, the quantum states 
which have the same density and phase profiles with those of a dark soliton 
have constructed \cite{Sato_2012,Sato_2016}; Then, it was found that the density profile of a single grey soliton appears in the quantum measurement process \cite{PhysRevA.92.032110(Syrwid-Sacha2015)}. Furthermore, the construction of quantum states of grey solitons has been elaborated with the Gaussian weights \cite{PhysRevA.99.043632(Shamailov-Brand2019),kaminishi2020construction}. 
In particular, an ansatz on how to assign  
the Gaussian weights to the yrast states 
was proposed \cite{kaminishi2020construction}. 
The matrix elements of the bosonic field operator of the Lieb-Liniger model are  characterized by grey solitons \cite{Sato_2016}, and double elliptic dark solitons are derived in association with the quantum measurement process \cite{PhysRevResearch.2.033368}.

In this article we demonstrate that quantum states characterized by a double dark or grey soliton are constructed in terms of the eigenstates with a pair of particle-hole excitations in the Lieb-Liniger model. It thus gives nontrivial examples where the description of the GP equation is effective 
in the quantum system.  
We shall also show that the ansatz of Ref.~\cite{kaminishi2020construction} is quite effective as far as the density profile is concerned.
The contents of the present article are given as follows.   
We first review  in \S 2.1 that quantum states associated with the density and phase profiles of single dark or grey solitons of the GP equation are constructed from the one-hole excitations, namely, the type II excitations in the Lieb-Liniger model. They are also called the yrast states. We then explain the ansatz~\cite{kaminishi2020construction} for constructing quantum dark solitons in terms of the ideal Gaussian weights in \S 2.2. We shall introduce the mean-field product state as the BEC state corresponding to a macroscopic wave function. Here we assume that the BEC state should exist for a given macroscopic wave function in the weak coupling regime of a finite system, if the coupling constant is small enough.  
 In \S 3.1  we introduce a novel quantum state which leads to an elliptic dark-soliton with a nonzero winding number in the phase profile of the matrix element of the bosonic field operator between the quantum states. In \S 3.2 we derive the elliptic multiple dark soliton solution of the GP equation.  We then argue analytically in  \S 3.3 that the phase profile should be consistent with that of the elliptic double dark-soliton in the limit of sending the system size to infinity.  
We present in \S 4 the density profiles of quantum double dark-solitons, explicitly. For an illustration  we discuss in \S 4.1 the results of the free fermions in one dimension. In \S 4.2 we show that the quantum state given by the sum of the two particle-hole states with an equal weight leads to the density and phase profiles, which 
almost perfectly agree with those of 
the elliptic double dark-soliton solutions of the GP equation, respectively, in the weak coupling regime. In \S 4.3, assigning the ideal Gaussian weights of the ansatz~\cite{kaminishi2020construction} 
to the quantum double dark-soliton states we obtain two distinct notches with different depths in the density profiles.
We also present the square amplitude and the phase profiles of the matrix element of the bosonic field operator between the quantum double dark soliton states with $N$ particles and $(N-1)$-particles. 
In the phase profiles of the quantum double dark soliton states we observe that one of the two notches has a nonzero winding number. 
Thus, the ansatz of the ideal Gaussian weights is also important 
for constructing double dark soliton states. 
However, we should point out that it does not completely lead to the mean-field product state. In fact, one of the key properties of the mean-field product state holds only approximately for all the examples investigated in this article.
On the other hand, we expect that one can improve the constructed quantum state with the ansatz so that one obtain a complete mean-field product state by increasing the number of particles $N$ and the system size $L$, while by decreasing the coupling constant $c$.

We now give some remarks on mathematical aspects. 
The derivation of soliton solutions is much harder in the repulsive case than in the attractive case for the NS (or GP) equation as far as theoretical formulation is concerned. A dark soliton solution of the GP equation was first derived by Tsuzuki for single solitons \cite{tsuzuki}. The multi-soliton solutions of the NS equation were constructed by Zakharov-Shabat, in particular, in the case of attractive nonlinear interaction \cite{zakharov}. For the repulsive case multi-soliton solutions were systematically derived by Faddeev and Takhtajan through the inverse scattering  method\cite{faddeev2007hamiltonian}. The multi dark soliton solutions are called ''soliton solutions in the case of finite density'', in the reference \cite{faddeev2007hamiltonian}. Here we should remark that the gap-soliton solutions of the NS equation were derived through the algebro-geometric approach, which generalize the  multi solitons of the NS equation into gap solitons\cite{belokolos1994algebro}.  The formulation of the inverse scattering method is quite general and useful,  and the algebro-geometric approach produces quite powerful results. However, both of them are mathematically rather rich and it seems that they are not easily accessible for standard theoretical physicists. Due to the mathematical richness they are not often referred to in the physics literature, in particular, in the context of cold atoms. It is thus one of the purposes of the present article as a review to point out some explicit connections of some researches in theoretical physics to the useful references in mathematics.  
 

\section{Quantum single dark soliton}

\subsection{Construction of quantum states of single dark solitons}\label{singledarksoliton}

In this section we review the method to construct quantum dark soliton states
in the case of a single soliton. 
Let us consider the one-dimensional Bose gas interacting via the $\delta$-function potential \cite{LL1963}
\begin{align}
\ham
= - \sum_{j=1}^{N} {\frac {\partial^2} {\partial x_j^2}}
+ 2c \sum_{j < k}^{N} \delta(x_j-x_k)
-\mu N,
\label{h1}
\end{align}
where 
$N$ is the number of particle, 
$c$ is the coupling constant and
$\mu$ is the chemical potential. 
We impose a periodic boundary condition with length $L$. 
In the second quantized form it reads
\begin{align}
    \ham
    = \int_{0}^{L} dx 
    \left[ 
    \partial_x \hat{ \psi}^{\dagger} \partial_x \hat{ \psi}
    + 
    c \hat{ \psi}^{\dagger} \hat{ \psi}^{\dagger} \hat{ \psi} \hat{ \psi}
    - 
    \mu \hat{ \psi}^{\dagger} \hat{ \psi} 
    \right],
\label{ham_LL}
\end{align}
where the operators $\hat{\psi}(x)$ and $\hat{\psi}^\dagger(x)$ are the Boson annihilation and creation operators satisfying the canonical commutation relations
\begin{align}
&[\hat{ \psi}(x), \hat{ \psi}^{\dagger}(x')]=\delta(x-x'), \\
&[\hat{ \psi}(x), \hat{ \psi}(x)]=[\hat{ \psi}^{\dagger}(x), \hat{ \psi}^{\dagger}(x)]=0. 
\end{align}

The exact eigenstates and eigenenergies  of this system are obtained through 
the Bethe ansatz method \cite{bethe1931, LL1963, L1963}. 
The central object of this method is the Bethe ansatz equation
\begin{align}
e^{i k_j L} = \prod_{\ell \ne j}^{N} 
\frac
{k_j - k_{\ell} + i c }
{k_j - k_{\ell} - i c }
\quad
\text{for}
\quad
j=1, 2, \ldots, N,
\label{BAE} 
\end{align}
the solution of which yields the exact wavefunction
\begin{align}
&\varphi_{k_1,\cdots,k_N}(x_1,\cdots,x_N)
=
\frac{c^{N/2}}{\sqrt{N!}}
\left( \prod_{j>\ell}^N\frac{1}{k_j-k_\ell} \right)
\sum_{\sigma\in S_N}^{N!}
A_\sigma
\exp\left[i\sum_{j=1}^N k_{\sigma_j} x_j\right], 
\\
&A_\sigma=(-1)^\sigma
\prod_{j>\ell}^N\left[k_{\sigma_j}-k_{\sigma_\ell}-i c \, \text{sign}(x_j-x_\ell)\right] 
\end{align}
with the energy $E$ and momentum $P$ eigenvalues
\begin{align}
    E=\sum_{j=1}^N k_j^2, \qquad
        P=\sum_{j=1}^N k_j. 
\end{align}
Recasting the Bethe ansatz equation \eqref{BAE} into the logarithmic form, one has
\begin{align}
k_j L = 2 \pi I_j - 2 \sum_{\ell \ne j}^{N} 
\arctan \left({\frac {k_j - k_{\ell}} c } \right)
\quad
\text{for}
\quad
j=1, 2, \ldots, N ,
\label{logBAE} 
\end{align}
where $I_j$'s are the Bethe quantum numbers, 
which are integers for odd $N$ and half-odd integers for even $N$.
In the case of repulsive interaction $c>0$, 
Bethe ansatz equation \eqref{logBAE} has a unique real solution 
for a given set of Bethe quantum numbers \cite{KorepinBook}. 
Especially the ground state is given by the set 
$I_j=-(N-1)/2+(j-1)$ for $j=1,2,\cdots,N$. 
Particle and hole excitations are constructed by putting particles and holes in this Bethe quantum numbers for the ground state. 

In the attractive case $c<0$, 
the mean field equation has the bright soliton solution 
and the corresponding quantum bound states are constructed in \cite{WadachiSakagami}. 
In this work we consider the repulsive case $c>0$, 
where the mean field equation has the dark soliton solution \cite{tsuzuki, zakharov}. 
In the refs. \cite{Sato_2012, Sato_2016}, it is shown that 
the quantum states corresponding to the dark soliton solutions are the superpositions of the hole excitations
\begin{align}
\ket{X(q), N} := \frac 1 {\sqrt{N}} \sum_{p=0}^{N-1} \exp(- 2 \pi i pq/N)  \ket{P(p), N} . 
\label{eq:XN}
\end{align}
Here $\ket{P(p), N}$ is the one-hole excitation with the total momentum $P(p)=2 \pi p/L$, whose quantum numbers are 
\begin{align} 
I_j & = -(N+1)/2+j \quad \mbox{for} \quad 1 \leq j \leq N-p \nonumber \\ 
 & = -(N+1)/2+j+1 \quad \mbox{for} \quad N-p+1\leq j \leq N . 
\label{eq:Ij-one-hole}
\end{align} 
The density profile of this state 
$\bra{ X, N}
\hat{ \psi}^{\dagger}(x) \hat{ \psi}(x)
\ket{X,N}$ 
shows a density notch at the position $X(q)=(q/N+1/2)L$, 
which coincides with the squared amplitude of the classical dark soliton solution \cite{Sato_2012, Sato_2016}. 
Here the matrix element
$\bra{ X, N}
\hat{ \psi}^{\dagger}(x) \hat{ \psi}(x)
\ket{X,N}$ 
can be effectively evaluated by the determinant formula
 for the norms of Bethe eigenstates \cite{Gaudin1983, Korepin1982} 
and that of the form factors of the density operator \cite{Slavnov1989, Slavnov1990, Calabrese2007} 
\begin{align}
&\bra {P',N}
\hat{ \psi}^{\dagger}(x) \hat{ \psi}(x)
\ket{P,N}
\nonumber\\&
=(-1)^{N(N+1)/2}(P-P')
\left(\prod^N_{j,\ell=1}\frac{1}{k'_j-k_\ell}\right) 
\left( \prod^N_{j>\ell}k_{j,\ell}k'_{j,\ell}\sqrt{\frac{{\hat K}(k'_{j,\ell})}{{\hat K}(k_{j,\ell})} } \right)
\frac{\det U(k,k')}{\sqrt{\det G(k)\det G(k')}}, 
\label{eq:Slavnov_density}
\end{align}
where the quasimomenta $\{k_1,\cdots,k_N\}$ and $\{k'_1,\cdots,k'_N\}$ 
give the eigenstates $\ket{P}$ and $\ket{P'}$, respectively. 
We use the abbreviations $k_{j,\ell}:=k_j-k_\ell$ and $k'_{j,\ell}:=k'_j-k'_\ell$. 
The kernel ${\hat K}(k)$ is defined by ${\hat K}(k)=2c/(k^2+c^2)$. 
The matrix $G(k)$ is called the Gaudin matrix, whose $(j,\ell)$ th element is given by 
\begin{equation} 
G(k)_{j,\ell}=\delta_{j,\ell}
\left[L+\sum_{m=1}^N {\hat K}(k_{j,m})\right]-{\hat K}(k_{j,\ell})
\quad \mbox{for} \quad j, \ell=1, 2, \cdots, N . \label{eq:Gaudin}
\end{equation} 
The matrix elements of the $(N-1)$ by $(N-1)$ matrix $U(k,k')$ are given by 
\begin{align}
U(k,k')_{j,\ell}=2\delta_{j\ell}\text{Im}
\left[\prod^N_{a=1}
\frac{k'_a-k_j + i c}{k_a-k_j + i c}
\right]
+\frac{\prod^N_{a=1}(k'_a-k_j)}{\prod^N_{a\neq j}(k_a-k_j)} 
\left({\hat K}(k_{j,\ell})-{\hat K}(k_{N,\ell})\right). 
\label{eq:matrixU_density}
\end{align}

We can also consider the matrix element of the single field operator
\begin{align}
\begin{split}
&
\bra{X', N-1} \hat{\psi}(x) \ket{X, N}
\\&
=
\frac1{\sqrt{N(N-1)}}
\sum^{N-1}_{p=0}
\sum^{N-2}_{p'=0}
\exp\left[2 \pi i (p-p')\frac{x}{L}\right]
\exp\left[-2 \pi i \left(\frac{pq}{N}-\frac{p'q'}{N}\right)\right]
\bra{P', N-1} \hat{\psi}(0) \ket{P, N},
\label{eq:matrix}
\end{split}
\end{align}
where $P=2\pi p/L$ and $P'=2\pi p'/L$ denote the total momenta 
of the normalized Bethe eigenstates in the type II branch 
$\ket{P, N}$ and $\ket{P', N}$, respectively. 
The determinant formula is given by
\cite{Gaudin1983, Korepin1982, Slavnov1989, Slavnov1990, Calabrese2007, Kojima1997}
\begin{align}
\begin{split}
&\bra{P', N-1} \hat{\psi}(0)\ket{P, N}
=(-1)^{N(N+1)/2+1}
\\&\times
\left(
\prod^{N-1}_{j=1}
\prod^N_{\ell=1}
\frac{1}{k'_j-k_\ell}\right) 
\left( 
\prod^N_{j>\ell}k_{j,\ell}\sqrt{k_{j,\ell}^2+c^2} 
\right)
\left( 
\prod^{N-1}_{j>\ell}\frac{k'_{j,\ell}}{\sqrt{(k'_{j,\ell})^2+c^2}} 
\right)
\frac{\det \widehat{U}(k,k')}{\sqrt{\det G(k)\det G(k')}}, 
\label{eq:Slavnov_field}
\end{split}
\end{align}
where the quasi-momenta $\{k_1,\cdots,k_N\}$ and $\{k'_1,\cdots,k'_{N-1}\}$ 
give the eigenstates $\ket{P, N}$ and $\ket{P', N-1}$, respectively. 
Here we remark that the symbol $X^{'}$ corresponds to an integer $q^{'}$.  
We recall that the matrix $G(k)$ denotes the Gaudin matrix, whose $(j,\ell)$th element is 
given in \eqref{eq:Gaudin}. 
%
%
The matrix elements of the $(N-1)$ by $(N-1)$ matrix $\widehat{U}(k,k')$ are given by 
\begin{align}
\widehat{U}(k,k')_{j,\ell}&=2\delta_{j\ell}\text{Im}
\left[
\frac{\prod^{N-1}_{a=1}(k'_a-k_j + ic)}{\prod^N_{a=1}(k_a-k_j + ic)}
\right]
+\frac{\prod^{N-1}_{a=1}(k'_a-k_j)}{\prod^N_{a\neq j}(k_a-k_j)}
\left({\hat K}(k_{j,\ell})-{\hat K}(k_{N,\ell})\right). 
\label{eq:matrixU_field}
\end{align}

\subsection{Construction with Gaussian weights}\label{subsec_singledarksoliton_g}

\subsubsection{Effective Bose-Einstein condensation in the weak coupling case}

In this subsection we briefly review the approach
~\cite{kaminishi2020construction} to quantum states with a dark soliton. We shall also investigate its validity numerically. In the approach a corresponding quantum state is defined by making use of a dark soliton solution  $\phi_P(x)$ of the classical nonlinear Schr\"odinger equation, and constructed explicitly with the Gaussian weights.

Based on the concept of the Bose-Einstein condensation (BEC) a ``single-particle dark soliton state''  is proposed as an ideal quantum state of a dark soliton in Ref.~\cite{kaminishi2020construction}:   
All the particles occupy the same one-particle state in the BEC. In the most strict sense, the BEC never happens in one dimension. However, in a weak-interaction regime of a large but finite system, 
a quasi-condensate state may appear. 
Thus, we assume a macroscopic wave function for the quantum dark-soliton state. 
Here we suggest that in a quasi-condensate state the coupling constant should be very small. It should be inversely proportional to the system size in order to keep the condensation fraction constant as the system size increases. The condensation fraction is evaluated in the ground-state of the Lieb-Liniger model for the repulsive coupling by applying the Onsager-Penrose criterion of the BEC~\cite{sato2013finitesize}.

The proposed BEC state ~\cite{kaminishi2020construction}, i.e., the quantum state introduced with the wave function $\phi_P$, which we shall call the mean-field product state, exactly reproduces the density and phase profiles of the classical dark-soliton $\phi_P$. 
Furthermore, it was numerically studied ~\cite{kaminishi2020construction} that the main components of the product state are given by a weighted sum of the yrast states. Here we remark that it is not trivial to express it in terms of the eigenstates of the Lieb-Liniger model.  

\subsubsection{Ansatz with the ideal Gaussian weights 
and the mean-field product state}

Let us explain the approach 
\cite{kaminishi2020construction}.
For a given macroscopic wave function  $\phi(x)$ and a given positive integer $N$ 
we consider a corresponding BEC state
consisting of $N$ particles. If all the particles occupy the same  single-particle state associated with the wave function $\phi(x)$,  
the $N$-particle state is given by 
$$\frac{1}{\sqrt{N!}}\left(\int_{-L/2}^{L/2}dx\,\phi(x)\hat{\psi}^\dagger(x)\right)^N\ket{\Omega},$$ where $\ket{\Omega}$ denotes the vacuum.
Here, the number $N$ may be macroscopically large. 
We call it the mean-field product state associated with the wave function $\phi(x)$ in the paper.
In Ref. ~\cite{kaminishi2020construction} 
as an $N$-particle quantum soliton state corresponding to a classical dark soliton solution $\phi_P(x-X)$ 
the following BEC state is proposed 
\begin{equation}
\ket{N,X;P}=\frac{1}{\sqrt{N!}}\left(\int_{-L/2}^{L/2}dx\,\frac{\phi_P(x-X)}{\sqrt{N}}\hat{\psi}^\dagger(x)\right)^N\ket{\Omega}.
\label{eq:Q_soliton}
\end{equation}
This quantum state exactly reproduces the classical dark soliton density profile which is given by
\begin{equation}
\bra{N,X;P}\hat{\psi}^\dagger(x)\hat{\psi}(x)\ket{N,X;P}_\mathrm{BEC}=|\phi_P(x-X)|^2,
\label{eq:becdensity}
\end{equation}
and satisfies
\begin{equation}
\hat{\psi}(x)\ket{N,X;P}=\phi_P(x-X)\ket{N-1,X;P}.
\label{eq:soliton_wave}
\end{equation}
Here we remark that an explicit analytic expression of a dark soliton solution  $\phi_P(x)$ in a finite system will be given in section 3.2~\cite{Sato_2016}.

\begin{figure}[ht!]
\begin{center}
\includegraphics[width=7.5cm]{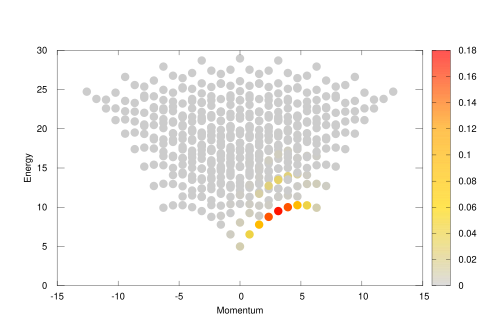}
\setlength\textfloatsep{7pt}
\vskip10mm
\caption{The overlap $|C_{\{I_j\}}|^2=|\braket{\{I_j\}_N|N,X;P=\pi}|^2$ for $n=1$, $c=0.1$, and $N=8$.
Each circle represents an eigenstate $\{I_j\}$ with the momentum $\sum_{j=1}^Nk_j$ and the energy $\sum_{j=1}^Nk_j^2$, and deep color implies a large overlap.
The eigenstate with the lowest energy for a given momentum $P$ corresponds to an yrast state $\ket{N,P}_\mathrm{yr}$.}
\label{fig:overlap}
\end{center}
\end{figure}

It should be nontrivial to express the mean-field product state 
$\ket{N,X;P}$ in terms of the energy eigenstates of the Lieb-Liniger model. However, an ansatz is given in Ref. ~\cite{kaminishi2020construction} that it should be given by the Gaussian weighted sum of yrast states.
Let us consider an expansion of the $N$-particle quantum soliton state $\ket{N,X;P}$ in terms of the eigenstates $\ket{\{I_j\}_N}$ of the Lieb-Liniger model as follows:
\begin{equation}
\ket{N,X;P}=\sum_{\{I_j\}}C_{\{I_j\}}\ket{\{I_j\}_N}.
\end{equation}
Quite an important numerical observation is made in  Ref.~\cite{kaminishi2020construction} that the  expansion coefficients of the yrast states
are dominant among all the coefficients. 
It is shown in Figure \ref{fig:overlap}. 
In other words, 
it is strongly suggested
that a quantum soliton state defined by Eq.~(\ref{eq:Q_soliton}) 
should be
essentially expressed as a superposition of the yrast states.

The expansion coefficients of the yrast states in the weak-coupling regime
are evaluated in Ref.~\cite{kaminishi2020construction} in the large $N$ limit. We now explain the method how to evaluate them.
In the thermodynamic limit with a fixed coupling constant $c$, the mean $P$ and the variance $\sigma_P^2$ of the total momentum operator $\hat{P}=\int_{-L/2}^{L/2}dx\,\hat{\psi}^\dagger(x)(-i\partial_x)\hat{\psi}(x)$ are given by
\begin{equation}
\lim_{N\to\infty}\braket{N,X;P|\hat{P}|N,X;P}=P
\label{eq: P_mean}
\end{equation}
and
\begin{equation}
\lim_{N\to\infty}\braket{N,X;P|(\hat{P}-P)^2|N,X;P}=\frac{4}{3}\beta^3 n \sqrt{n c} \equiv \sigma_P^2,
\label{eq:P_variance}
\end{equation}
respectively. Here $n$ denotes the density of particles: $n=N/L$ and $\beta$ 
will be introduced in section \ref{subsub:reduction} as  $\beta=1-v^2/v^2_{c, \infty}$. 
We remark that $n$ and $\beta$ correspond to $\rho_0$ and $\gamma$ in Ref. \cite{kaminishi2020construction}.
Associated with Eqs. \eqref{eq: P_mean} and \eqref{eq:P_variance} we assume  that in the BEC state $\ket{N,X;P}$ the momenta $\{p_j\}_{j=1}^N$ of the $N$ particles can be regarded as independent random variables. Hence, the central limit theorem implies that the total  momentum $\sum_{j=1}^Np_j$ obeys a Gaussian  distribution. It thus follows from the above assumption that if the quantum soliton state consists of only the yrast states, the former is given by a Gaussian superposition of the latter for any large system size:
\begin{align}
\ket{N,X;P}&\approx\mathcal{N}^{-1/2}\sum_{P'}e^{iP'(X-L/2)}e^{-G_P(P')/2}\ket{N,P'}_\mathrm{yr}
\nonumber \\
&\equiv\ket{N,X;P}_\mathrm{yr},
\label{eq:Q_soliton_yrast}
\end{align}
where 
\begin{equation}
G_P(P')=\frac{(P'-P)^2}{2\sigma_P^2} 
\label{eq:Gaussian_weights}
\end{equation}
and $\mathcal{N}$ is a normalization constant, and the sum is taken over $P'=2\pi M/L$ with $M=0,1,\dots,N$.

\begin{figure*}[h!]
\begin{minipage}[b]{0.32\linewidth}
  \centering
  \includegraphics[width=5.0cm]{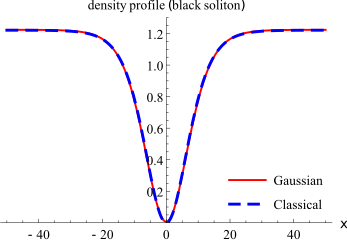}\\
 \end{minipage}
 \begin{minipage}[b]{0.32\linewidth}
  \centering\bigskip
  \includegraphics[width=5.0cm]{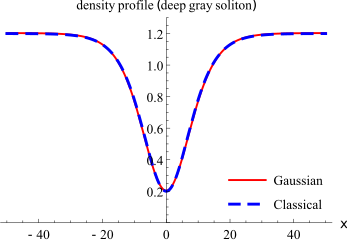}\\
 \end{minipage}
 \begin{minipage}[b]{0.32\linewidth}
  \centering
\includegraphics[width=5.0cm]{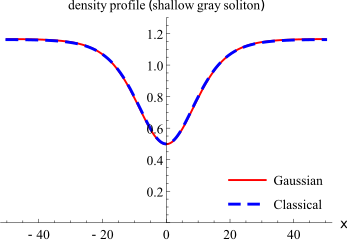}
 \end{minipage}
\caption{Comparison of the density profiles of quantum soliton states $\ket{N,0;P}_\mathrm{yr}$ (the Gaussian superposition of yrast states) and those of classical dark solitons,
for $P=\pi n$ (left), $P=0.4502 \pi n$ (middle) and $P=(\pi/2-1) n$ (right). The system size is set as $N=L=100$.}
\label{fig:density}
\end{figure*}

In Fig.~\ref{fig:density} it is clear that  
the density profiles of the quantum and classical solitons completely agree with each other.
Here we remark that the density profiles for the quantum soliton state of Eq.~(\ref{eq:Q_soliton_yrast}),   $\braket{N,0,P|\hat{\psi}^\dagger(x)\hat{\psi}(x)|N,0,P}$ with $P=\pi n$ (the black soliton) and $P=(\pi/2-1) n$ (a gray soliton) 
are plotted for $N=L=100$ and $c=0.01$ together with the corresponding classical solutions of the GP equation in Fig.~\ref{fig:density}. 
Furthermore, it is also clear that the expression of  Eq.~(\ref{eq:Q_soliton_yrast}) is much more useful for numerical calculation 
than Eq.~(\ref{eq:Q_soliton}). The range of the summation in Eq.~(\ref{eq:Q_soliton_yrast}) is  restricted only to the yrast states and the expression of expansion coefficients is explicitly given. 
Thus, we can calculate almost any physical observables by making use of the Bethe ansatz method even for very large system sizes. 

As far as the density profiles are concerned 
we might conclude
that the quantum soliton state should be
completely constructed by the superposition of the yrast states with the ideal Gaussian weights in Eq.~(\ref{eq:Q_soliton_yrast}) with Eq.~(\ref{eq:Gaussian_weights}). 
In fact, the ansatz with the ideal Gaussian weights~\cite{kaminishi2020construction} has the following two important properties: 
(i) We can construct a family of quantum dark solitons by changing the value of $P$, where $P$ determines the depth of the soliton. This is definitely an advantage of Eq.~(\ref{eq:Q_soliton_yrast}) compared to the uniform superposition introduced in Ref.~\cite{Sato_2012} because the uniform superposition of the yrast states cannot control the depth of the soliton. Figure~\ref{fig:density} shows that the Gaussian superposition indeed reproduces dark solitons with an  arbitrary depth (i.e., an arbitrary $P$), while the uniform superposition does not;  
(ii) The Gaussian weights in Eqs.~(\ref{eq:Q_soliton_yrast}) and (\ref{eq:Gaussian_weights}) assigned to the yrast states are completely determined without any free parameter. Here we recall that the Gaussian form is derived by applying the central limit theorem to the quantum soliton state in Eq.~(\ref{eq:Q_soliton}) and the analytic expression of $\sigma_P$ is obtained. 
Here it should be remarked that the idea of a Gaussian superposition of yrast states has also been proposed in a recent work~\cite{PhysRevA.99.043632(Shamailov-Brand2019)}, where the Gaussian form is assumed and the Gaussian width $\sigma_P$ is given by a fitting parameter.

\subsubsection{Numerical study on the validity of the ansatz with the ideal Gaussian weights}

\begin{figure}[H]
    \centering
    \includegraphics[width=13cm]{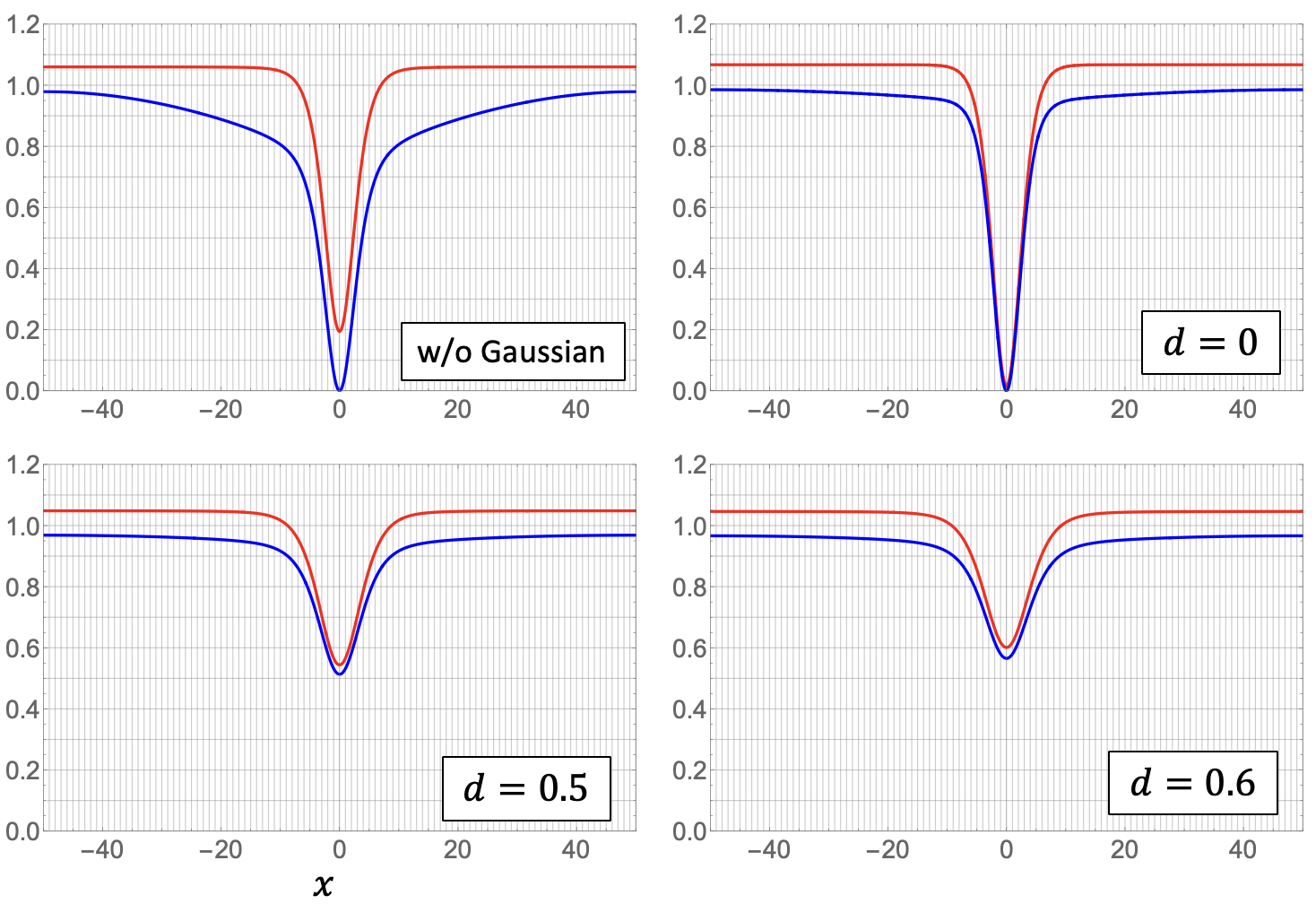}
    \caption{Density profiles (red curves) and the profiles of the square amplitude of the matrix element of the field operator between the quantum soliton states with 
    $N$ and $N-1$ particles (blue curves) for $N=L=100$ and the coupling constant $c=0.1$. The vertical axis represents the local density. The profiles of the quantum soliton state without Gaussian weights are plotted in the upper left panel. In other panels, the profiles for the quantum soliton states with the ideal Gaussian weights are plotted for three values of target soliton depth $d$ as 
    $d=0, 0.5$, and $0.6$, respectively. They   correspond to the parameters of the Gaussian weights, $(P, \sigma^2)=(\pi, 0.421637), (0.18169\pi, 0.149071)$, and $(0.124027\pi, 0.106667)$, respectively, through Eqs. \eqref{eq:g_param_p} and \eqref{eq:g_param_sigsq}.}
    \label{fig:g_den_amp}
\end{figure}

The ansatz with the ideal Gaussian weights~\cite{kaminishi2020construction} should thus be quite useful and significant with respect to the density profile. However, we now point out that it does not completely produce the mean-field product state and also that  
Eq.~(\ref{eq:soliton_wave}) does not hold exactly but only approximately.  
In Figure \ref{fig:g_den_amp} the density profiles (red curves) are plotted together with the corresponding profiles (blue curves) of the square amplitude of the matrix element of the field operator between the quantum soliton states with $N$ and $N-1$  particles. The two plots do not completely overlap: The former plots are larger than the latter plots at any position of $x$. Here they are plotted for $N=L=100$ and $c=0.1$. It is the case for all the examples investigated so far. 
In the three panels of Figure \ref{fig:g_den_amp}
the density and square-amplitude profiles are plotted 
for the quantum soliton states constructed by the ansatz of the ideal Gaussian weights for three different values of soliton momentum $P$. 
Here we remark that the soliton momentum $P$ and the variance $\sigma_P^2$ are determined by the target soliton depth $d$ through Eqs. \eqref{eq:g_param_p} 
and \eqref{eq:g_param_sigsq}. We also remark that the definition of target soliton depth will be given in Eq. \eqref{eq: target_depth}.

For each value of the three values of depth $d$ the density profile of the quantum soliton state constructed by the ansatz is clearly larger than the corresponding square-amplitude profile. 
Here we remark that the degree of overlap of the two profiles suggests the degree how far the quantum soliton state constructed by the ansatz of the ideal Gaussian weights~\cite{kaminishi2020construction} is consistent with the mean-field product state, as we shall argue in section \ref{sec:difference}.

We suggest that the quantum states constructed by the ansatz of the ideal Gaussian weights should be much closer to the mean-field product state than those of the superposition of the yrast states with equal weight \eqref{eq:XN}. 
In Figure \ref{fig:g_den_amp}
we observe that in each value of target soliton depth $d$ which will be defined by Eq. \eqref{eq: target_depth} 
the difference of the two profiles is smaller than that of the profiles of the quantum states without the Gaussian weights. 
Furthermore, we expect that one can improve the quantum soliton state constructed by the ansatz so that we make it closer to an ideal mean-field product state, by increasing the number of particles $N$ and the system size $L$, while by decreasing the coupling constant $c$.  

%
%
\section{Quantum state of a dark soliton with nonzero winding number and the corresponding classical solution}

We present the quantum state associated with an elliptic dark-soliton solution of the GP equation with a nonzero winding number, in this section. Hereafter, we call it the quantum state of a dark soliton with a nonzero winding number. In order to define the winding number we briefly construct elliptic multiple dark-solitons of GP equation and derive the single dark-soliton in the thermodynamic limit. 

\subsection{Novel quantum state of a single dark soliton with a nonzero winding number}

\subsubsection{Construction of quantum states with nonzero winding number}

There exists a classical periodic solution of the NS equation $\phi(x) = \sqrt{\rho(z)}\exp[i\varphi(x)]$ satisfying the periodic boundary conditions 
\begin{align}
    \varphi(x+L) = \varphi(x) + 2\pi J
    \label{eq:pbc_winding}
\end{align}
where $J$ is an arbitrary integer. It is called the phase-winding number \cite{PhysRevA.79.063616}. 
We construct the quantum state of a single 
dark-soliton with a nonzero winding number. When we construct the quantum dark-soliton state as a superposition of Bethe eigenstates, we select a different sequence of Bethe eigenstates, which have not been investigated in the previous studies \cite{Sato_2016, kaminishi2020construction}. 
Let us consider a new quantum state in the following:  
\begin{align}
\ket{X(q), N}_{J=1} := \frac 1 {\sqrt{N}} \sum_{p=1}^{N} \exp(- 2 \pi i pq/N)  \ket{P(p), N}. 
\label{eq:newXN}
\end{align}
The difference between the quantum states \eqref{eq:XN} and \eqref{eq:newXN} is only one point that the lower and upper bounds are shifted by 1 in the summation with respect to $p$. That is, the summation of Eq. \eqref{eq:newXN} is obtained from that of Eq. \eqref{eq:XN} by removing the ground state ($p=0$) and adding the quantum state with momentum $P=2\pi/L\ (p=N)$. 

We express the matrix element of the field operator $\hat{\psi}(x)$ for a quantum single dark soliton state $\ket{X, N}$ with a non-zero winding number by
\begin{align}
    \psi_Q(x) :=  [\braket{X,N-1|\hat{\psi}(x)|X,N}]_{J=1}
    \label{eq:matele_expression}
\end{align}
and write $\psi_Q(x)$ in the polar form: $\psi_Q(x)=|\psi_Q(x)|\exp[{i\arg[\psi_Q(x)]}]$. The matrix element is computed by the determinant formula  such as Eq. \eqref{eq:Slavnov_field}.

\subsubsection{Profiles of the square amplitude and the phase of the matrix element of the field operator}

\begin{figure}[H]
    \centering
    \includegraphics[width=13cm]{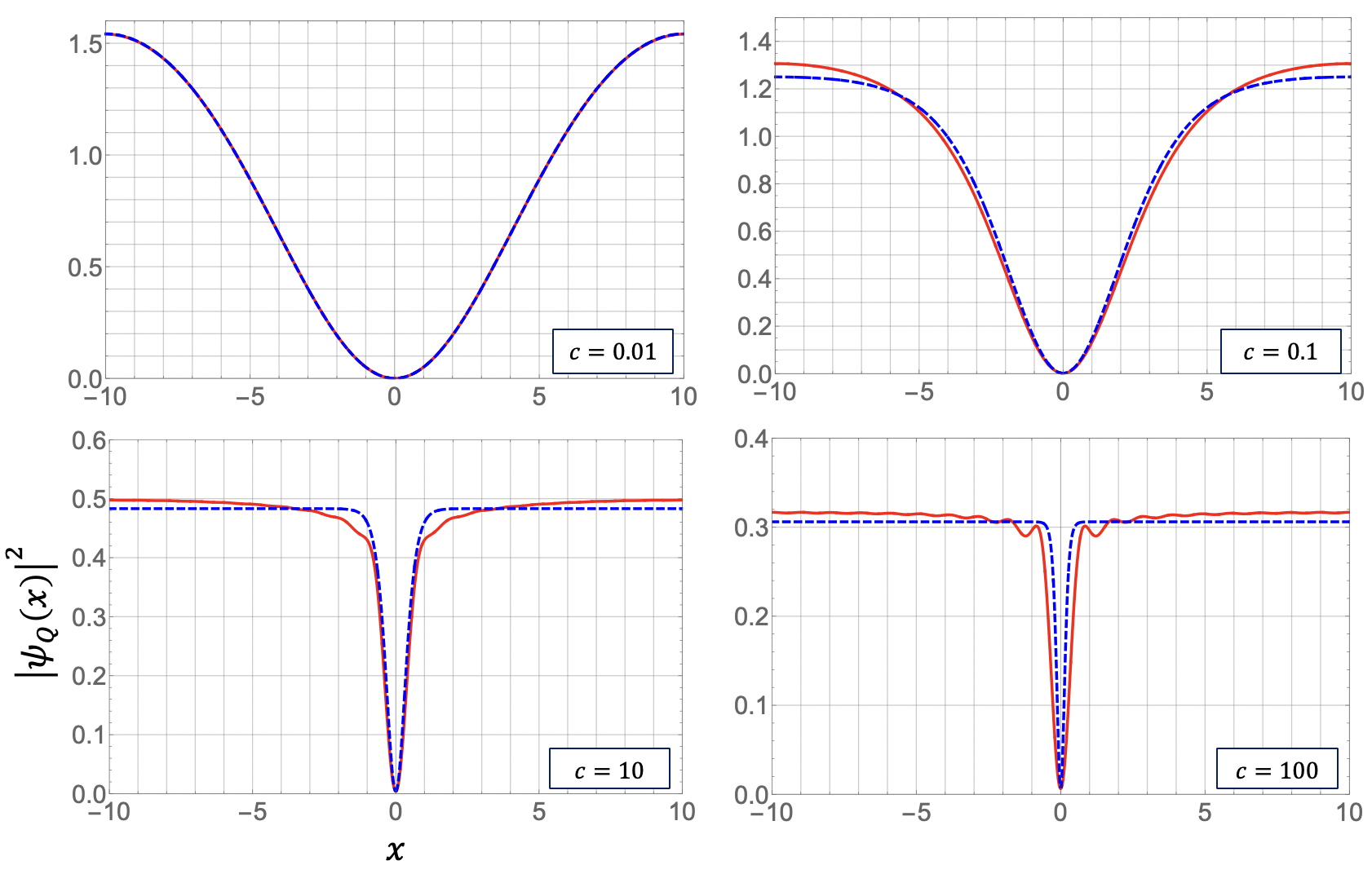}
    \caption{Square amplitudes of the matrix elements $|\psi_Q(x)|^2$ for $N=L=20,$ and $c=0.01,\ 0.1,\ 10,\ \text{and}\ 100$ (red solid curves) and the corresponding elliptic solutions (blue dashed curves). The corresponding quantum state is given by Eq. \eqref{eq:newXN}, which has $J=1$. In the panel of $c=0.01$, the red solid curve agrees with the blue dashed curve.}
    \label{fig:sq_amp_winding}
\end{figure}

\begin{figure}[H]
    \centering
    \includegraphics[width=13cm]{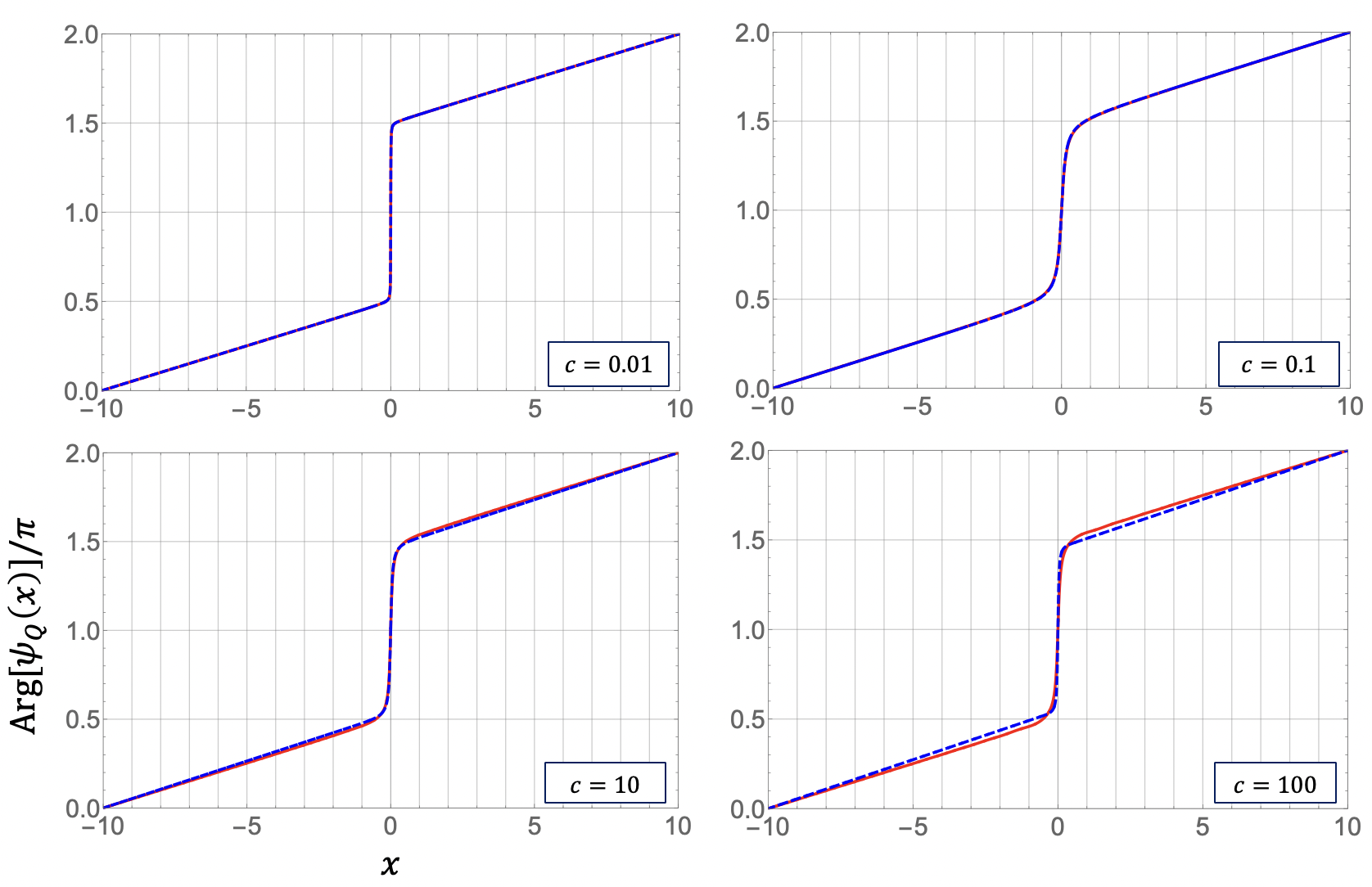}
    \caption{Phase profiles $\arg[\psi_g(x)]/\pi$ for $N=L=20,$ and $c=0.05,\ 0.1,\ 10,\ \text{and}\ 100$ (red solid curves) and the corresponding elliptic solutions (blue dashed curves). The corresponding quantum state is given by Eq. \eqref{eq:newXN}, which has $J=1$. In the panels of $c=0.01$ and 0.1, the red solid curves completely agree with the blue dashed curves, respectively.}
    \label{fig:arg_winding}
\end{figure}

Figures \ref{fig:sq_amp_winding} and \ref{fig:arg_winding} present the profiles of the square amplitude $|\psi_Q(x)|^2$ and those of the phase field $\arg[\psi_Q(x)]$ in red solid curves, respectively, for four different values of the coupling constant such as $c=0.01, 0.1, 10$ and 100. The plots of the corresponding elliptic solutions of the GP equation are exhibited in blue dashed curves for the four values of the coupling constant $c$. Here the system size $L$ is given by $L=20$ and the number $N$ of particles by $N=20$, so that the density $n$ is given by 1: $n=N/L=1$.   

When we search the corresponding classical solution of the GP equation to the density profile of a given quantum dark-soliton state, we numerically evaluate the soliton depth $d$ in the density profile, and then determine the value of modulus $k$ through the relation: $a_1 =d$, as will be shown in section \ref{sec:chemical_potential}. Here $a_1$ denotes the smallest root of the equation: $U(\rho)=0$, where the potential $U(\rho)$ will be given by Eq. \eqref{eq:U}. 
We can draw the density profile of the corresponding elliptic dark-soliton by applying Eq. (\ref{eq:rho-EK}), as will be shown in section 
\ref{subsec_ellipticmulti}. Here we remark that an elliptic dark-soliton has four parameters $c$, $L$, density $n$ and modulus $k$. We calculate the chemical potential $\mu$ and the critical velocity $v_c$ by Eqs. (\ref{eq:mu-k}) and (\ref{eq:critical_velocity}), respectively, as will be shown later. Moreover, when we search the corresponding classical elliptic dark-soliton solution to the matrix element of the field operator for a given quantum dark-soliton state, we first evaluate the density $n$ by numerically integrating the whole profile of the square amplitude of the matrix element. We secondlly evaluate the soliton depth $d$ by finding out the minimum of the square amplitude profile. We then determine the value of modulus $k$ through the relation: $a_1 =d$.


\begin{table}[ht]
\centering
\caption{Parameters of the profiles of the elliptic solutions in Figures \ref{fig:sq_amp_winding} and \ref{fig:arg_winding}. The notch depth $d$ is given by the minimum of the square amplitude profile for each value of $c$. }
\begin{tabular}[t]{ccccccc}
\hline
$c$ & $n$ & $d$ & $k$ & $\mu/2n c$ & $v_c/2\sqrt{nc}$ & $v/v_c$  \\
\hline
0.01  & 0.830 840 & 9.670 81 $\times10^{-7}$ & 0.680 75 & 1.446 70 & 1.753 96 &  0.982 297\\
0.1 & 0.903 971 & 0.00172 103 & 0.993 356 & 1.272 54 & 0.853 159 & 0.575 645\\
1.0 & 0.788 041 & 0.00192 928 & 1- 5.693 69$\times10^{-8}$& 1.111 45 & 0.899 118 & 0.147 633\\
10 & 0.461 185 & 0.00419 681 & 1- 7.864 09$\times10^{-19}$ & 1.05 184 & 0.949 564 & -0.0188 539\\
100 & 0.300 536 & 0.00640 668 & 1- 2.2907 21$\times10^{-47}$ & 1.022 293 & 0.977 873 & -0.117 299\\
\hline
\end{tabular}
\label{tab:param_winding_matele}
\end{table}%

In Figure \ref{fig:sq_amp_winding}, the plots of the square amplitudes $|\psi_Q(x)|^2$ agree with those of the corresponding elliptic solution represented with the blue dashed lines, in particular, for the weak coupling cases of $c=0.01$ and 0.1. In the cases of $c=0.01, 0.1$ and 10,  the width of the notch decreases in the square amplitude profile as the coupling constant increases. In the case of $c=100$, 
however, the notch of the square amplitude $|\psi_Q(x)|^2$ is wider than that of the elliptic solution, as shown in Figure \ref{fig:sq_amp_winding}.

In Figure \ref{fig:arg_winding} all 
the phase profiles $\arg[\psi_Q(x)]$ exhibited 
satisfy the boundary condition
\begin{align}
    \arg[\psi_Q(x+L)] = \arg[\psi_Q(x)] +2\pi \, . 
\end{align}
It corresponds to the case of $J=1$ in Eq. \eqref{eq:pbc_winding}. Moreover, the phase profiles  of the quantum states almost perfectly agree with those of the classical solutions. 
All the blue dotted curves for the phase profiles of the classical solutions with $J=1$ are depicted by choosing the plus in the plus/minus sign in the constant $W$, which will be defined in Eq. \eqref{eq:1st-integral1}. The choice should be consistent with the abruptly increasing behavior appearing at each notch position, as we shall argue in section  \ref{sec:abrupt_change}. 

The parameters for the fitted curves in 
Figures \ref{fig:sq_amp_winding} and \ref{fig:arg_winding} are listed in Table \ref{tab:param_winding_matele}. 
The estimates of the velocity parameter $v$ are given by making use of Eq. (\ref{eq:velocity} ) 
with $J=1$.  
For the strong coupling cases of $c=10$ and $100$ 
the velocity parameter $v$ becomes negative, which will be explained with Eqs. \eqref{eq:vJ=1} and (\ref{eq:condition_positive}) in section  \ref{sec:nonzero_winding_number}.

\begin{figure}[H]
    \centering
    \includegraphics[width=13cm]{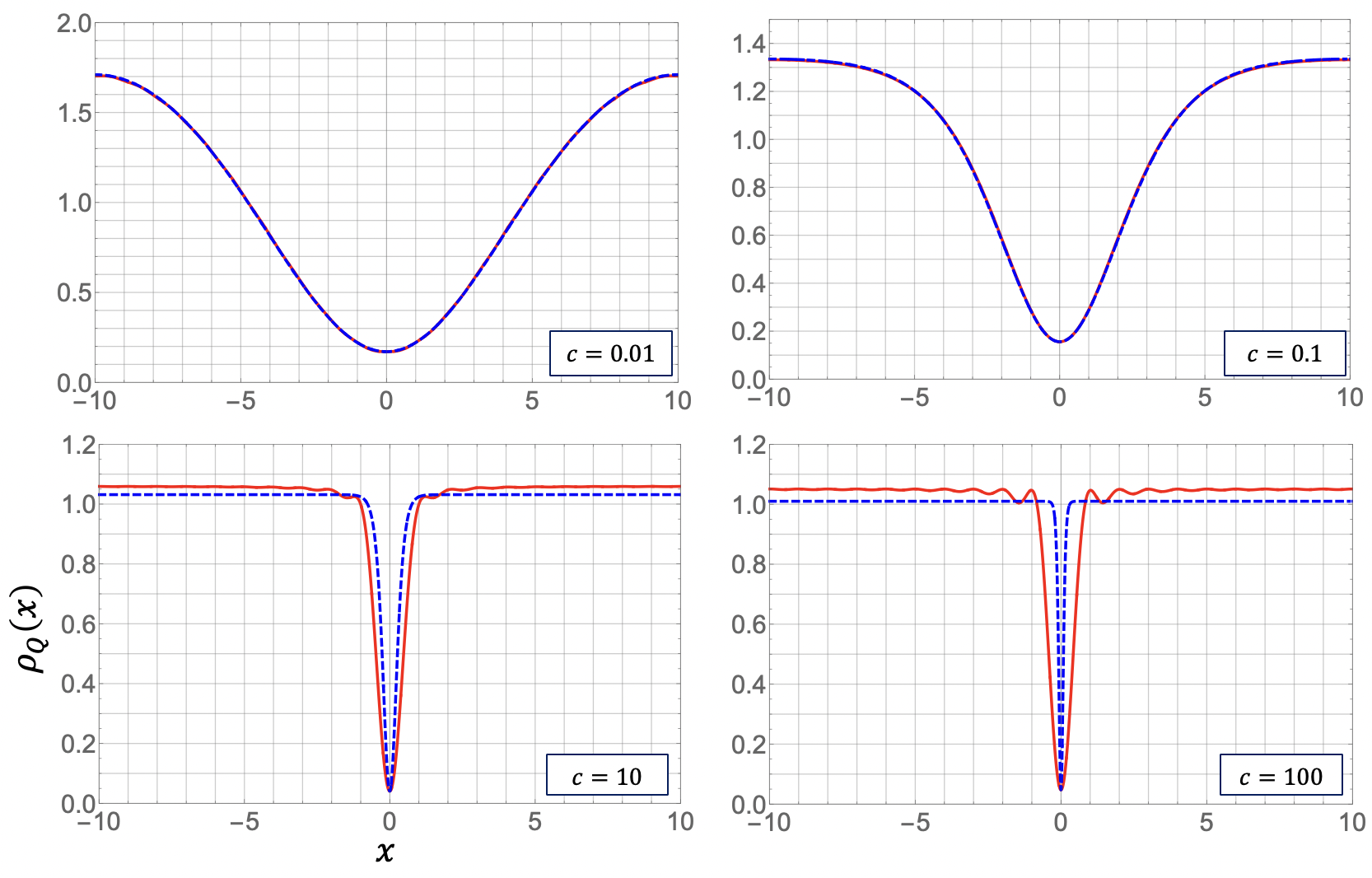}
    \caption{Density profiles for $N=L=20,$ and $c=0.01,\ 0.1,\ 10,\ \text{and}\ 100$ (red solid curves), and the corresponding elliptic solutions (blue dashed curves). The corresponding quantum state is given by Eq. \eqref{eq:newXN}, which has $J=1$. In the panels of $c=0.01$ and 0.1, the red solid curves completely agree with the blue dashed curves, respectively.}
    \label{fig:den_winding}
\end{figure}

\begin{table}[ht]
\centering
\caption{Parameters of the density profiles of the elliptic dark-soliton solutions in Figure \ref{fig:den_winding}. 
}
\begin{tabular}[t]{cccccc}
\hline
$c$ &  $d$ & $k$ & $\mu/2n c$ & $v_c/2\sqrt{nc}$ & $v/v_c$  \\
\hline
0.01 & 0.170 659 & 0.680 343 & 1.690 17 & 1.444 67 &  0.978 349\\
0.1  & 0.155 884 & 0.991 939 & 1.40 035 &  0.667 869 & 0.317 074\\
10 & 0.0406 314 &1- 3.64 668$\times10^{-17}$ & 1.04 120 & 0.959 205 & -0.151 773\\
100 & 0.0475 028 & 1- 4.973 27$\times10^{-85}$ & 1.013 54 & 0.986 482 & -0.202 798\\
\hline
\end{tabular}
\label{tab:param_winding_den}
\end{table}%

In Figure \ref{fig:den_winding} the density profiles $\left[\braket{X(q), N|\hat{\psi}^\dagger(x)\hat{\psi}(x)|X(q), N}\right]_{J=1}$ of the novel quantum state \eqref{eq:newXN} are depicted 
for the four values of the coupling constant $c=0.01, 0.1, 10$ and 100 with red solid curves.
The blue dotted curves are the density profiles 
of the corresponding classical solutions, i.e., 
elliptic single dark-solitons. Here we recall that they are depicted by evaluating only the soliton depth $d$.

It is interesting to observe that the density profiles of the quantum state with $J=1$ in Figure \ref{fig:den_winding} are similar to those of $J=0$ depicted in earlier studies. 
It is consistent with the fact that the local density $\rho(x)$ of the classical solution does not depend on the winding number $J$.

The parameters for the fitted curves in 
Figure \ref{fig:den_winding} are listed in Table \ref{tab:param_winding_den}. The estimates of the velocity parameter $v$ are given by making use of Eq. (\ref{eq:velocity}) with $J=1$. Similarly as in Table 
\ref{tab:param_winding_matele},  
for the strong coupling cases of $c=10$ and $100$ 
the velocity parameter $v$ are negative. It  will be explained with Eqs. \eqref{eq:vJ=1} and (\ref{eq:condition_positive}) in section  \ref{sec:nonzero_winding_number}.

Finally we remark that quantum states of a single dark soliton with zero winding number, i.e., $J=0$, 
have been studied in the previous study
 \cite{Sato_2016}. In all the phase profiles the argument $\arg[\psi_Q(x)]$ satisfies the periodic boundary condition: $\arg[\psi_Q(x)] = \arg[\psi_Q(x+L)]$.



\subsection{Construction of an elliptic multiple dark soliton for the GP equation}
\label{subsec_ellipticmulti}

\subsubsection{Local density \texorpdfstring{$\rho(x)$}{} and the phase field  \texorpdfstring{$\varphi(x)$}{} of an elliptic multiple dark-soliton}





Let us consider stationary solutions of the GP equation in the region of length $L$ under the periodic boundary conditions. 
We assume that they depend on the coordinate $x$ and time $t$ through variable $z=x-vt$ as follows:
\begin{equation}
\psi(x,t) = \phi (x-vt) . 
\end{equation}
Substituting it into Eq. (\ref{eq:GPE}) we have 
\begin{equation}
\phi^{''} - i v \phi^{'} + \mu \phi - 2c |\phi|^2 \phi = 0.  \label{eq:f}
\end{equation}
If there are $N$ particles in the system of length $L$, the average density $n$ is given by $n=N/L$.
Here we remark that in the limit of sending 
$L$ to infinity, the chemical potential $\mu$ approaches $2 c n$ if the local density  $|\phi(x)|^2$ approaches $n$ for some large region. We express it as $\mu_{\infty}=2 c n$. 

We express the complex scalar field $\phi(z)$ in terms of the local density $\rho(z)$ and the phase field $\varphi(z)$ as follows.  
\begin{equation}
\phi(z) = \sqrt{\rho(z)} \exp\left( i \varphi(z) \right).  \label{eq:ansatz_esol}
\end{equation}
By putting the ansatz (\ref{eq:ansatz_esol}) into Eq. (\ref{eq:f}), from the real part and the imaginary part of the derived equation we can show that there are two first-integrals of motion:  
\begin{equation} 
    \frac {d \varphi}{dx} = \frac v 2 + \frac W {\rho(x)} ,  \label{eq:1st-integral1}
\end{equation}
and 
\begin{equation}
\left( \frac{1}{2} \frac{d \rho}{dx} \right)^2 + U(\rho)  =  0 , \label{eq:1st-integral2}
\end{equation}
where 
\begin{equation}
    U(\rho)= -c \rho^3 + \left( \mu + {\frac {v^2} 4} \right) \rho^2 - 2V \rho + W^2. \label{eq:U} 
\end{equation}
Here $W$ and $V$ are constants with respect to the first integrals (\ref{eq:1st-integral1}) and (\ref{eq:1st-integral2}), respectively. 

Let $a_1$, $a_2$ and $a_3$ denote the three roots of the equation: $U(\rho)=0$. 
We assume that they are real. We put them in increasing order: $a_1 \le a_2 \le a_3$.  
It follows from the first integral (\ref{eq:1st-integral2}) 
that they are non-negative. The coefficient of $\rho^3$ in the potential $U(\rho)$ of Eq. (\ref{eq:U}) is negative since $c$ is positive: $c>0$, and hence the potential $U(\rho)$ is negative or zero only in the regions $a_1 \le \rho \le a_2$ and $a_3 \le \rho$. Physically 
the local density $\rho$ should be  non-negative and have an upper bound. It thus follows that we consider only the region $a_1 \le \rho \le a_2$ and $a_1$ is non-negative: $a_1 \ge 0$. In fact, the potential $U(\rho)$ is non-negative at the origin: $U(\rho=0)= W^2 \ge 0$. %
It is straightforward to show the following: 
\begin{equation}
    W = \pm \sqrt{c a_1 a_2 a_3}. 
    \label{eq:W} 
\end{equation}

By solving Eq. (\ref{eq:1st-integral2}) 
we express the local density $\rho(z)$ in terms of the Jacobi's elliptic function ${\rm sn}(u, k)$. 
Integrating the differential $2 dx = d \rho / \sqrt{-U(\rho)}$ with the initial condition: $\rho(x=0)=a_1$ and applying the variable transformation: $\rho= a_1 + (a_2-a_1)z^2$, we have 
\begin{equation}
\rho(x)= a_1 - (a_1-a_2) {\rm sn} ^2 \left( \sqrt{c(a_3-a_1)} \, x , k \right) \, ,  
\end{equation} 
where $k$ denotes the elliptic modulus  
\begin{equation}
    k = \sqrt{\frac {a_3- a_1} {a_2 - a_1}} . 
    \label{eq:modulus}
\end{equation}
We now assume the period $L/m$ with a positive integer $m$ for stationary solutions. we thus have  
\begin{equation}
\sqrt{ c (a_3- a_1) } = {2 m K}/L ,  \label{eq:K/L}
\end{equation}
where $K$ denotes the complete elliptic integrals of the first kind. 


We now express the three roots $a_1, a_2$ and $a_3$ in terms of the average density $n$, $c$, $L$ and $k$. Since the average of the local density over the period $L$ is equal to the average density $n$, we have  
\begin{equation}
\frac 1 L \int_{0}^{L} \rho(x) dx = n . 
\label{eq:average-density}
\end{equation}
Through Eq. (\ref{eq:modulus}) and the relation: ${\rm sn}^2(u,k) + k^2 {\rm dn}^2(u,k)=1$,  
the local density $\rho(z)$ is expressed in terms of the Jacobi's elliptic function ${\rm dn}(u,k)$ as  
\begin{equation}
    \rho(x) = a_3 - (a_3-a_1) {\rm dn}^2(2 m K x/L, \, k). 
\end{equation}
It follows from the definition of the complete elliptic integral of the second kind: $E=\int_0^K {\rm dn}^2(u, \, k)du$ and Eq. (\ref{eq:average-density}) that we have 
\begin{equation}
    a_3 = n + \frac 1 c \left( \frac {2 m K} L  \right)^2 \, \frac E K. 
\end{equation}
Let us introduce a parameter $\beta_k$ by 
\begin{equation} 
\beta_k = \frac 1 {n c} \left( \frac {2 m K} L \right)^2 \, . 
\end{equation}
It follows from Eqs. (\ref{eq:modulus}) and (\ref{eq:average-density}) that the three roots are expressed as 
\begin{eqnarray}
a_1 & = & n \left(1 + \beta_k  \left( \frac E K -1 \right)  \right) \, , \nonumber \\ 
a_2 & = & n \left(1 + \beta_k  \left( \frac E K -1 + k^2 \right)  \right) \, , \nonumber \\ 
a_3 & = & n \left(1 + \beta_k \frac E K \right)
\, .
\end{eqnarray}
Thus, the three roots $a_1, a_2, a_3$ are expressed in terms of $\beta_k$, the density $n$, and the modulus $k$. The parameter  $\beta_k$ depends on $c$,  $n$, $k$  and $L/m$.  Here the integer $m$ denotes the number of notches in the multiple dark-soliton, i.e., the multiplicity.  
The parameter $\beta_k$  has to satisfy the condition: $\beta_k \le 1/(1-E/K)$ since $a_1$ is non-negative: $a_1 \ge 0$.  

The local density is expressed in the form of the average density $n$ and the oscillating part  
\begin{equation}
\rho(x)= n + 
n \, \beta_k \, 
\left( \frac E K - {\rm dn} ^2 \left( \frac {2m K x} L , k \right) \right) \, .  
\label{eq:rho-EK}
\end{equation} 
Here we remark that the second term in the right-hand side of Eq.  \eqref{eq:rho-EK} leads to a logarithmic correction associated with the conservation of the number of particles when we send $k$ to 1 and $L$ to infinity  with $K/L$ fixed. 

Let us now express the phase field in terms of elliptic functions. By integrating Eq. (\ref{eq:1st-integral1}) with respect to variable $x$ we have 
\begin{equation}
    \varphi(x)-\varphi(0) 
    = {\frac v 2 } x + {\frac W {a_1}} 
    \int_{0}^{x} 
    \frac {dy} {1- n_e \, {\rm sn}^2\left({2m K y}/L , k \right) }
\end{equation}
where $n_e$ is given by  
\begin{equation}
    n_e 
  = 1 - a_2/a_1 \, . 
\end{equation}
In terms of  the elliptic integral of the third kind $\Pi(r, \phi, k)$ 
with $\sin \phi = {\rm sn}(u,k)$ and a given number $r$ 
\begin{equation}
\Pi(r, \phi={\rm am}\, u, k) 
= \int_{0}^{u} \frac {dv} {1 - r \, {\rm sn}^2(v, k)}        
\end{equation}
we have 
\begin{equation}
    \varphi(x)-\varphi(0) 
    = {\frac v 2 } x + {\frac W {a_1}} 
{\frac L {2m}} {\frac 1 K} 
\Pi\left(n_e, \phi={\rm am}\, \frac {2mK x} L, k\right).    
\label{eq:phase}
\end{equation}

In order to determine the velocity $v$, 
let us consider the shifted periodic boundary condition (\ref{eq:pbc_winding}) with a winding number $J$. 
\begin{equation}
\varphi(x=\frac L {2m}) - \varphi(x=0) 
= {\frac  {2 \pi J} {2m}} .
\label{eq:phase_bc}
\end{equation}
Putting Eq. (\ref{eq:phase}) into it,  
we express the velocity $v$ in terms of elliptic functions as follows: 
\begin{equation}
    v  = - \frac {2 W} {a_1} \frac 1 K \Pi(n_e, k) + {\frac {4 \pi} L } J.
    \label{eq:velocity} 
\end{equation}

We now show that the phase field $\varphi(x)$ is expressed in terms of Jacobi's Theta function.
Let $\Theta(u)$ and $Z(u)$ denote Jacobi's Theta and Zeta functions, respectively.   
 Making use of the connection of the Theta functions to the elliptic integral of the third kind \cite{copson}:
\begin{equation}
    \int_{0}^{u} \frac {k^2 {\rm sn}(a, k) {\rm cn}(a, k) {\rm dn}(a, k) {\rm sn}^2(u, k)} 
    { 1- k^2 {\rm sn}^2(a, k) {\rm sn}^2(u, k) } 
    \, du 
= \frac 1 2 \log \left( \frac {\Theta(u-a)}{\Theta(u+a)} \right) + u Z(a) 
\end{equation}
we derive the following formula
\begin{eqnarray}
\int_0^u \frac {dv} {1-r {\rm sn}^2(v, k) } 
& = & u \left(1 + Z(a) \sqrt{\frac {r} {(k^2-r) (1-r)}}\right) + \nonumber \\
& & \qquad + \frac 1 2 \log \left( \frac {\Theta(u-a)}{\Theta(u+a)} \right) 
\sqrt{\frac {r} {(k^2-r) (1-r)}} \label{eq:formula-EI3}
\end{eqnarray}
where $r$ is any real number while  
the parameter $a$ is related to $r$ by 
$k^2 {\rm sn}^2(a,k) =r$. 
In order to apply Eq. (\ref{eq:formula-EI3}) to 
the phase field, we set $r= n_e=1-a_2/a_1$. We define the parameter $a$ by a pure imaginary number $a= i \alpha$ with $\alpha > 0$ satisfying the following relation: 
\begin{equation} 
 {\rm sn}(i \alpha, k) = 
i \sqrt{\frac {\beta_k} {1 - \beta_k (1-E/K)} } \, . \label{eq:alpha} 
\end{equation} 
Then, we can numerically evaluate the real number $\alpha$ by making use of the relations: ${\rm sn} u = \Theta(K) H(u) /H(K) \Theta(u)$  and $\sqrt{k}= H(K)/\Theta(K)$. 

Thus, making use of the relation between the elliptic integral of the third kind and 
Jacobi's Zeta and Theta functions, we express the phase field $\varphi(x)$ as the logarithm of a ratio of Theta functions 
\begin{equation}
\varphi(x) -\varphi(0) = {\frac {2 \pi J} L} x  \pm  {\frac i 2} 
\log \left(
\frac {\Theta ( {2m K x}/L  - i \alpha ) } 
                    {\Theta ( {2m K x}/L + i \alpha ) }  \right) \, ,     
\label{eq:phase-theta}
\end{equation} 
for $W= \pm \sqrt{c a_1 a_2 a_3}$. Here the plus/minus signs are in correspondence.  
In the derivation of Eq. (\ref{eq:phase-theta}), through Eq. (\ref{eq:formula-EI3}) 
we have expressed the complete elliptic integral of the third kind in terms of the Zeta function
\begin{equation}
    \Pi(r, k) = K \left(1 + 
    Z(a, k) \sqrt{\frac r {(k^2-r) (1-r)}}
    \right) .  \label{eq:CEI3-Zeta}
\end{equation}

\subsubsection{Chemical potential and the critical velocity}
\label{sec:chemical_potential}
When the system size $L$, the coupling constant $c$, and the density $n$ are fixed, once the modulus $k$ is specified, the corresponding classical solution of the multiple dark-soliton with multiplicity $m$ is determined. Here, the chemical potential $\mu$ is related to the parameters $c$, $n$, $L$ and $k$ as follows: 
\begin{equation}
    \mu = 3 n c  - {\frac {n c} 4 } v^2 + 
    {\frac 4 {\left( L/m \right)^2}} \, K (3E - (2-k^2) K). 
\label{eq:mu-k}
\end{equation}
Here we remark that velocity $v$ is given by Eq. (\ref{eq:velocity}) with a winding number $J$. 

For a given quantum dark-soliton state, 
we search the corresponding elliptic solution of the GP equation, as follows. Suppose that 
 the system size $L$, the coupling constant $c$, and the density $n$ are also given. We then determine the modulus $k$ by equating the first root $a_1$ with the depth $d$ at the notch, i.e., the minimum value of the density profile. 
 Furthermore, we evaluate the chemical potential $\mu$ by Eq. (\ref{eq:mu-k}). Here we recall that velocity $v$ is evaluated by Eq. \eqref{eq:velocity}. 

Let us derive the range of velocity $v$. 
When the system size $L$, the coupling constant $c$, and the density $n$ are fixed, and the chemical potential $\mu$ is also fixed, then the value of velocity $v$ specified by Eq. (\ref{eq:mu-k}) is a function of modulus $k$. We thus have the range of velocity $v$ as we change the value of modulus $k$ from 0 to 1. 
We can show that the function $f(k)=K (3E - (2-k^2) K)$ is monotonically decreasing with respect to modulus $k$. The largest value of $f(k)$ is therefore given by $f(0)=\pi^2/4$.
It follows that the largest value  $v_c$ of the velocity in the range is given by 
\begin{equation}
    v_c = \sqrt{(2 m \pi/L)^2 + 4(3nc-\mu)}. 
    \label{eq:critical_velocity}
\end{equation}
We call it the critical velocity.

\subsection{Reduction of elliptic multiple dark-soliton}

\subsubsection{Reduction of an elliptic multiple dark soliton into a series of single dark-solitons}
\label{subsub:reduction}

We shall show that an elliptic multiple dark-soliton 
reduces into separate single dark-solitons if we 
increase the system size in a certain way. 
Let us consider the limit of sending the system size $L$ to infinity ($L \to \infty$) while the modulus $k$ to 1 ($k \to 1$) simultaneously 
so that the ratio $K(k)/L$ is kept constant. 
We define the ratio $b$ by $b=K/L$. 
Here we remark that we keep the density $n$ constant. 
We define $\beta$ by the simultaneous limit of sending $L$ to infinity and $k$ to 1 with $K/L=b$: 
\begin{equation} 
\beta = \lim_{k \to 1, L \to \infty} \beta_{k} = \frac {(2m b)^2} {c n} \, . 
\end{equation}   

We introduce the complementary modulus $k^{'}$ by $k^{'}= \sqrt{1-k^2}$. 
We define $K^{'}(k)$ and $E^{'}(k)$ by $K^{'}(k)= K(k^{'})$ and $E^{'}(k)= E(k^{'})$, 
respectively. We remark that $K$ and $i K^{'}$ are the  quarter-periods of Jacobi's elliptic functions. 
We define an expansion parameter $p$ by 
\begin{equation}
p= \exp(-\pi K/K^{'}),     
\end{equation}
which is small when modulus $k$ is close to 1. 


We now consider the density profile of the stationary solution of an elliptic $m$-fold multiple dark-soliton at time $t=0$. We assume that at time $t=0$
the position of the $j$-th notch $x_j$ is given by $x_j= j L/m$ for $j = 0, 1, \ldots, m-1$,  in the interval $[0, L]$. Near the position of the $j$th notch, we express the $x$-coordinate in terms of the position of the $j$-th notch and the deviation from it:  
\begin{equation}
    x= \Delta x + x_j. 
\end{equation}
Here we assume that $|\Delta x|$ is small: $|\Delta x| \ll L/2m$.  
Near the $j$-th notch the square amplitude $\rho(x)$ is expanded 
as an asymptotic series of  parameter $p$ as follows: 
\begin{equation} 
\rho(x)/n = 1 - \beta \, \text{sech}^2 ( 2 b \, \Delta x ) + 2 \beta/\log(1/p) + \cdots \, . 
\end{equation} 
It is easy to show that the asymptotic expansion of the square amplitude 
up to the order of $1/\log p$, i.e., up to the order of $1/L$, satisfies the normalization condition. 
Thus, near the $j$-th notch, the density profile  of the elliptic multiple dark-soliton reduces to a single dark-soliton in the limit of sending $L$ to infinity and the elliptic modulus $k$ to 1 with the ratio $K/L$ fixed as $K/L=b$: 
\begin{equation}
\rho(x)=\left(\frac {\mu_{\infty} } {2c} \right) \left\{ 1 -\beta \, {\rm sech}^2 \left[ 
\left( \frac  {\beta\mu_{\infty} } 2 \right)^{1/2} \, ( x - x_j) \right] 
\right\} .  \label{eq:j-th-amp}
\end{equation}
Here $\beta$ is expressed as 
\begin{equation}
\beta=1-\left( \frac v {v_{c, \infty}} \right)^2    
\end{equation}
where $v_{c, \infty} = \sqrt{2 \mu_{\infty}}$. Here we remark that we have derived the square amplitude of the single dark-soliton due to Tsuzuki \cite{tsuzuki} 
in the limit of sending $p$ to zero.
Hereafter we express $\mu_{\infty}$ simply as $\mu$ for convenience.

We now consider the phase profile near a notch of an elliptic multiple dark-soliton. We recall that near the $j$th notch at time $t=0$ we have expressed the $x$-coordinate as $x=\Delta x + x_j$ where $x_j=jL/m$. 
In the limit of sending $L$ to infinity and 
$k$ to 1 with $K/L=b$ fixed  
the phase field is expressed in terms of the logarithmic functions as follows:
 \begin{eqnarray}
 \varphi(x) - \varphi(0) 
 & = & {\frac {2 \pi J} L} \, x \,   
 \mp \frac 1 {2 i}  \log 
 \left( 
 1 - i \sqrt{\frac {\beta} {1-\beta}} 
 \tanh\left(\sqrt{\frac {\mu \beta} 2 }\, \Delta x \right)\right)
\nonumber \\ 
&  & \qquad \pm \frac 1 {2 i}  \log 
\left(1 + i \sqrt{\frac {\beta}{1- \beta}} 
\tanh\left( \sqrt{\frac {\mu} {2} \beta} \, \Delta x \right) \right) \quad + 
O(1/\log(1/p)) \label{eq:phase-field}
\end{eqnarray}
for $W= \pm \sqrt{c a_1 a_2 a_3}$. Here 
the plus/minus signs are in correspondence.

For every $j$-th notch we have the expression (\ref{eq:j-th-amp})
of the local density as a function of the deviation $\Delta x$
from the $j$-th notch at time $t=0$. Thus, in the limit of sending $L$ to infinity 
and $k$ to 1 where ratio $K/L$ is fixed as $K/L=b$ 
an elliptic multiple dark-soliton reduces into a series of separated single dark-solitons.

\subsubsection{Abrupt change of the phase profile near a notch}
\label{sec:abrupt_change}

We express the position $x$ as $x=\Delta x + x_j$ near the $j$th notch at time $t=0$. It follows from Eq. (\ref{eq:phase-field}) that for very small $|\Delta x|$ the phase field $\varphi(x)$ is approximately given by the following if the system size $L$ is very large 
\begin{equation}
\varphi(x)  - \varphi(0)
=  {\frac {2 \pi J} L} \left(\Delta x + x_j \right)
\pm \sqrt{\frac {\beta} {1 - \beta} } \,  
\sqrt{\frac {\mu \beta} 2} \, \Delta x  + o(|\Delta x|) \quad (|\Delta x| \ll 1).
\end{equation}
Thus, if $L$ is very large, 
the phase change near the $j$-th notch is approximately given by 
\begin{equation}
\varphi(x)  - \varphi(0) 
 - {\frac {2 \pi J} L}
\, x_j \approx 
\left( \pm \sqrt{\frac {\mu (1-\beta)} {2} }  \cdot \frac {\beta} {1-\beta}
+ {\frac {2 \pi J} L}
\right) \, \Delta x 
\quad (|\Delta x| \ll 1) 
\label{eq:abrupt} 
\end{equation}
for $W= \pm \sqrt{c a_1 a_2 a_3}$. 
Here we recall that the plus/minus signs are in correspondence.

The gradient of the phase near the $j$-th notch is therefore approximated by 
\begin{equation}
    \left. \frac {d \varphi}{dx}\right|_{\Delta x=0} = 
     \pm \sqrt{\frac {\mu (1-\beta)} {2} }
    {\frac {\beta} {1-\beta}}
    + {\frac {2 \pi J} L}
    \label{eq:phi-gradient}
\end{equation}
for $W = \pm \sqrt{c a_1 a_2 a_3}$. 
Thus, the phase field $\varphi(z)$ may abruptly increase 
($W>0$) or decrease ($W<0$) near the $j$th notch $x_j=jL/m$ 
if $|J|$ is not large, in particular, when $\beta$ is close to 1, since $1/\sqrt{1-\beta}$ can be much larger than 1. 

When the winding number is zero: $J=0$, the velocity parameter $v$ is positive if $W$ is negative from Eq. \eqref{eq:velocity} 
or as we shall show in Eq. \eqref{eq:v-zeta}.  The phase field abruptly decreases at the $j$th notch, i.e., we have the minus sign in Eq. (\ref{eq:abrupt}).   


The abrupt change of the phase at the notches of multiple dark-soliton will appear in the phase profile of quantum states corresponding to double dark-solitons in \S 4.

\subsubsection{Evaluation of velocity in an infinite limit} 

We first recall that through Eq. (\ref{eq:CEI3-Zeta}) the velocity is expressed in terms of the Zeta function:
\begin{equation}
    \frac v 2  
     =  - \frac W {a_1} \, 
    \left(1 + Z(a, k) \sqrt{\frac {n_e} {(k^2-{n_e}) (1-{n_e})}} \right) 
+ {\frac {2 \pi J} L} \,   \label{eq:v-zeta}
\end{equation}
where $W$ is given by $W = \pm \sqrt{c a_1 a_2 a_3}$. 

Let us evaluate the velocity $v$ in the limit of sending $L$ to infinity and 
$k$ to 1 where $K/L$ is kept constant as $K/L=b$. 
Through the Jacobi's imaginary transformation we can show 
\begin{equation}
    Z(i \alpha, k) = i \sqrt{\frac {\beta} {1-\beta} } + O(1/\log(1/p)) 
\quad (p \rightarrow 0). \label{eq:zeta}
\end{equation}
Putting it into Eq. (\ref{eq:v-zeta})
we have 
\begin{equation}
    \frac v 2 \approx \mp \sqrt{\frac {\mu (1-\beta)} {2} } + {\frac {2 \pi J} L}
\qquad (p \ll 1) 
\end{equation}
for $W = \pm \sqrt{c a_1 a_2 a_3}$. 

We recall that for $J=0$ the absolute value of the phase gradient at a  
notch is given by 
\begin{equation}
    \left. \frac {d \varphi}{dx}\right|_{\Delta x=0} = 
    {\frac v 2} \cdot {\frac {\beta} {1-\beta}} .   \label{eq:v-phi}
\end{equation}
It is larger than the value of velocity $v$ if $\beta$ is close to 1. Here we remark that the factor $\beta/(1-\beta)$ is larger than 1 if $\beta > 1/2$. 

\subsubsection{Velocity for a nonzero winding number}
\label{sec:nonzero_winding_number}

Let us consider the velocity $v$ in the case of $J=1$. For $W=\pm \sqrt{c a_1 a_2 a_3}$ we have the 
following approximation for $v$ if the system size $L$ is very large 
\begin{equation}
    \frac v 2 = \mp \sqrt{\frac {\mu (1-\beta)} {2}} + \frac {2 \pi} L.  
    \label{eq:vJ=1}
\end{equation}
When $W>0$, the condition that $v$ is positive is given by  the following:
\begin{equation}
    c n (1- \beta) < \left( \frac {2 \pi} L \right)^2. \label{eq:condition_positive}
\end{equation}
It holds if the coupling constant $c$ is small or $\beta$ is close to 1.

If the system size $L$ is very large, 
the phase change near the $j$th notch of an elliptic $m$-fold multiple dark-soliton 
is approximately given by 
\begin{equation}
    \varphi(x) - \varphi(0) - 
    \frac {2 \pi}{ L } x_j \approx 
    \left( \pm \sqrt{\frac {\mu (1-\beta)}{ 2} } \,  
\frac {\beta} {1-\beta} \,  
+ {\frac {2 \pi} L}
\right) \, \Delta x 
\qquad (|\Delta x| \ll 1) \label{eq:DeltaPhiJ=1}
\end{equation}
for $W=\pm \sqrt{c a_1 a_2 a_3}$.

It follows from Eqs. (\ref{eq:vJ=1}) and (\ref{eq:DeltaPhiJ=1}) 
that if $\beta$ is close to 1 it happens very rarely that both the phase gradient $d \varphi/dx$ and the velocity $v$ are positive simultaneously. It occurs only if the coupling constant $c$ is very small. 

For instance, if $W$ is negative, the sign in Eq. (\ref{eq:vJ=1}) is positive and hence $v$ is positive, while that 
of Eq. (\ref{eq:DeltaPhiJ=1}) is negative. 
If $\beta$ is very close to 1 then the factor $1/(1-\beta)$ is large, 
so that the phase gradient is negative unless
the coupling constant $c$ is extremely small such as it is proportional to $1/L^2$.  

On the other hand, if $W$ is positive, the sign in Eq. (\ref{eq:vJ=1}) is negative and hence $v$ is positive only if the inequality in Eq.  (\ref{eq:condition_positive}) holds, 
while that of Eq. (\ref{eq:DeltaPhiJ=1}) is positive and hence the phase gradient is positive. Thus, for  $W >0$,   both the phase gradient $d \varphi/dx$ and the velocity $v$ are positive simultaneously if the inequality   (\ref{eq:condition_positive}) holds.

\section{Quantum double dark soliton}
\subsection{Density profile in the free fermions}\label{freefermi}
\begin{figure}[H]
    \centering
    \includegraphics[width=15cm]{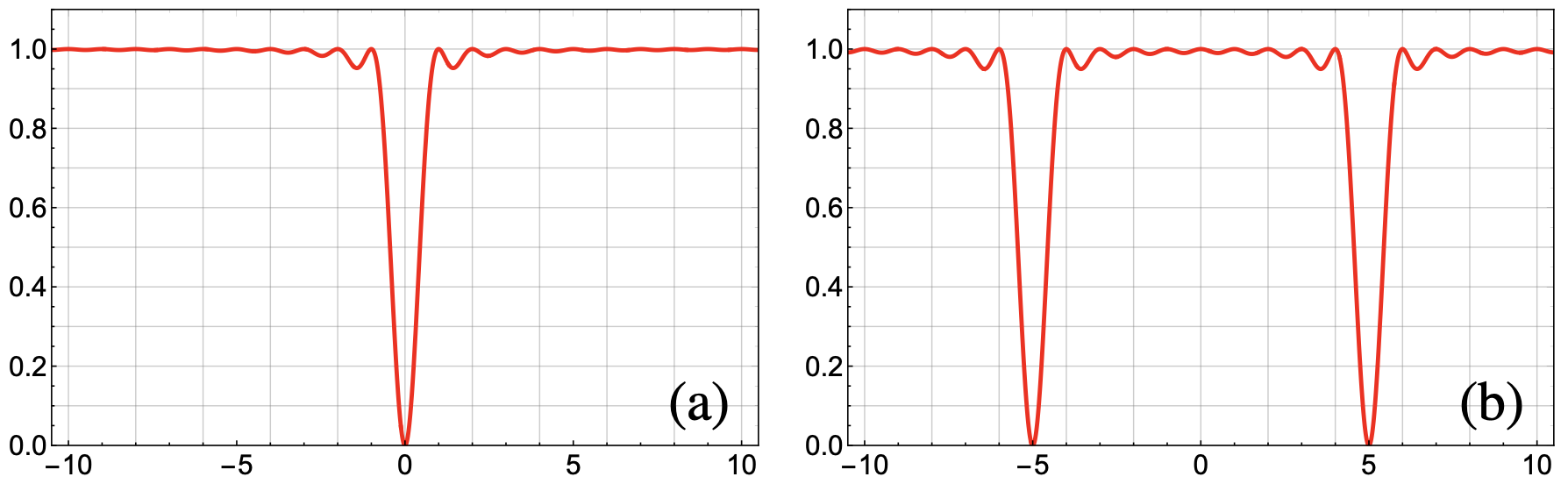}
    \caption{Density profiles in the free fermions. (a) Density profile of the state of Eq. \eqref{eq:fermion_onehole} with $N=L=21,\ $ and $X_1 = 0$. (b) Density profile of the state of Eq. \eqref{eq:fermion_twoholes} with $N=L=21,\ X_1 = -5,\ $and $X_2 = 5$. }
    \label{fig:fermidptwo}
\end{figure}

In order to explore a quantum state in the Lieb-Liniger model whose density profile has two notches, we now consider the one-dimensional free fermionic system, which corresponds to the Lieb-Liniger model in the limit of sending the coupling constant to infinity \cite{Girardeau1960}. First, we construct the quantum state in the free fermions which possesses one notch in its density profile shown in Figure \ref{fig:fermidptwo} (a). Then, we derive a quantum state whose density profile has two notches. 

Let us introduce the second quantized free fermions under the periodic boundary conditions with period $L$. The creation (annihilation) operator $a_\ell^\dagger$ ($a_\ell$) creates (annihilates) a particle with a momentum $k_\ell=2\pi n_\ell/L$, where $n_\ell$ is an integer. The operators $a_\ell$ and $a_\ell^\dagger$ satisfy the anti-commutation relations
\begin{align}
    &\{ a_\ell, a^\dagger_m\} = \delta_{\ell, m}, 
    &\{ a_\ell, a_m\} =  \{a^\dagger_\ell, a^\dagger_m\}=0.
\end{align}
The field operators for the one dimensional free fermions $\psi(x)$ and $\psi^\dagger(x)$ are defined by the creation and annihilation operators respectively
\begin{align}
   &\psi(x)=\frac{1}{\sqrt{L}} \sum_{\alpha=1}^{\infty} \mathrm{e}^{\mathrm{i} k_{\alpha} x} a_{\alpha}, 
   &\psi^{\dagger}(x)=\frac{1}{\sqrt{L}} \sum_{\alpha=1}^{\infty} \mathrm{e}^{-\mathrm{i} k_{\alpha} x} a_{\alpha}^{\dagger}.
\end{align}

We consider a quantum state possessing one notch in its density profile
\begin{align}
\ket{\Phi_{1}} &=\psi(X_1)\ket{gs} 
=\frac{1}{\sqrt{\mathcal{N}_1}} \sum_{\alpha=1}^{N}(-1)^{\alpha} e^{i k_{\alpha} X_1} \prod_{\ell=1 ; \ell \neq \alpha}^{N} a_{\ell}^{\dagger}\ket{0}.
\label{eq:fermion_onehole}
\end{align}
It is obtained by applying a field operator $\psi(X_1)$ at the position $x=X_1$ to the $N$-particle ground state $\ket{gs}$ defined by
\begin{align}
    \ket{gs} \equiv \prod_{\ell=1 }^{N} a_{\ell}^{\dagger}\ket{0}.
\end{align}
Here, $\mathcal{N}_1$ in Eq. \eqref{eq:fermion_onehole} is the normalization factor.
The state $\ket{\Phi_{1}}$ is the superposition of such states obtained by removing a particle with momentum $k = 2\pi n_\alpha /L$ from the $N$-particle ground state where the index $\alpha$ runs from 1 to $N$. The phase factor $e^{-ik_\alpha X_1}$ in Eq. (\ref{eq:fermion_onehole}) has a position $X_1$, implying the position of the notch in the expectation value of the local density operator for its state \cite{Sato_2016}.

Generalizing the state $\ket{\Phi_1}$, we assume that the fermionic state having two notches is a state such that the ground state with two field operators at $x= X_1,\ X_2$, that is,
\begin{align}
\ket{\Phi_2}
=\psi(X_2)\psi(X_1)\ket{gs}
= \frac{1}{\sqrt{\mathcal{N}_2}}  \sum_{\alpha<\beta}^{N}(-1)^{\alpha+\beta}\left(e^{i\left(k_{\alpha} X_{1}+k_{\beta} X_{2}\right)}-e^{i\left(k_{\beta} X_{1}+k_{\alpha} X_{2}\right)}\right) \times \prod_{\ell=1 ;\ \ell \neq \alpha, \beta}^{N} a_{\ell}^{\dagger}|0\rangle.
\label{eq:fermion_twoholes}
\end{align} The state $\ket{\Phi_2}$ is constructed by the double summation of the $N$-particle states which are given by removing two particles momenta $k_\alpha=2\pi n_\alpha/L$ and $k_\beta=2\pi n_\beta/L$ in the ground state. Here, the sum is taken over $\alpha$ and $\beta$. Figure \ref{fig:fermidptwo} (b) shows the density profile for the state (\ref{eq:fermion_twoholes}). It has two notches at the positions $x = X_1 $ and  $X_2$.

\subsection{Quantum double dark soliton with equal weight} \label{subsec_qdoublestate}
\subsubsection{Construction of the quantum state for a double dark soliton with equal weights}
By analogy with the fermionic case in the previous section, we now construct quantum double dark solitons associated with a superposition of two-hole excitation states in the Lieb-Liniger model. The choice of the Bethe quantum numbers is the key to construct the quantum double dark soliton state.

\begin{figure}[H]
    \centering
    \includegraphics[width=5cm]{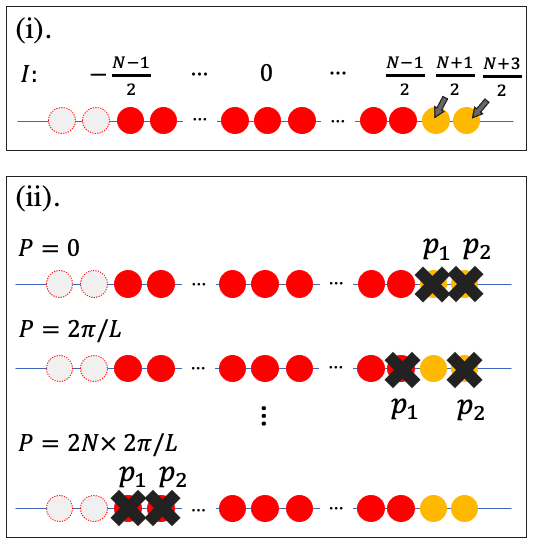}
    \caption{Configurations of the Bethe quantum numbers. (i) Configuration $I^+$, which is obtained by adding two particles (yellow circles) to the ground state (red circles).  (ii) Two holes, which correspond to a subset $\bm{p}=\{p_1,p_2\}$ of $I^+$, are punched in the Bethe quantum numbers $I^+$ with the value of $P$ defined in \eqref{total_momentum}. }
    \label{fig:bnum_den}
\end{figure}

A quantum state for a double dark soliton for $N$ particles is given by
\begin{align}
\ket{X_1,X_2,N}=\frac{1}{\sqrt{\mathcal{M}}}\sum_{\bm{p}\in \bm{P}}e^{i(p_1X_1+p_2X_2)}\ket{p_1,p_2,N}
\label{eq:2quantum_soliton_state}
\end{align}
with the normalization constant $\mathcal{M}$, and the two holes' positions $\bm{p}:=\{p_1, p_2\}$ in the set $\bm{P}$ indicating all possible two holes' positions for the set of Bethe quantum numbers $I^+$. The set $I^+$ consists of the ground state of $N$ particles and additional two particles, as shown in Figure \ref{fig:bnum_den}. For example, in the case of $N=5$, $I^+$ is the set $\{-2,-1,0,1,2,3,4\}$, and $\bm{P} = \left\lbrace\{-2,-1\},\{-2,0\},\{-2,1\},\{-2,2\},\cdots, \{2,4\}, \{3,4\}\right\rbrace$, which has $|\bm{P}|=\binom{7}{2}=21$ elements. The quantum double dark soliton state is a superposition of $|\bm{P}|=\binom{N+2}{2}$ two-hole excitation states.



With respect to the state \eqref{eq:2quantum_soliton_state}, the quantum double dark soliton with two notches at $X_1$ and $X_2$ is expressed by the expectation value of the local density operator
\begin{align}
\begin{split}
    \rho_Q(x) := &\braket{X_1,X_2,N|\hat{\rho}(x)|X_1,X_2,N}\\
	=&\frac{1}{\mathcal{M}}\sum_{\bm{p},\bm{p'}\in \bm{P}} e^{i(P-P')x}e^ {i(p_1X_1+p_2X_2)}e^ {-i(p'_1X_1+p'_2X_2)}\braket{p_1',p_2',N|\hat{\rho}(0,0)|p_1,p_2,N},
	\label{eq:2quantum_soliton}
\end{split}
\end{align}
where $P$ is the total momentum of the state with the system length $L$, which is given by
\begin{align}
P=\frac{2\pi}{L}\sum_{m\in I}m,\quad I \equiv I^+-\bm{p}.
\label{total_momentum}
\end{align}
Again, the form factors $\braket{P_1,P_2,N|\hat{\rho}(0,0)|P_1',P_2',N}$ are computed by the determinant formula. We recall that the notch positions of the quantum dark soliton are at $x=X_1$ and $X_2$, as in the case of the quantum single dark solitons constructed by  quantum states such as $\ket{X(q) ,N}$ in section \ref{singledarksoliton} and $\ket{N,X;P}$ in section \ref{subsec_singledarksoliton_g}, both of which have a notch at the position $X$.

\subsubsection{Density profile of the quantum double dark soliton state}

\begin{figure}[H]
    \centering
    \includegraphics[width=13cm]{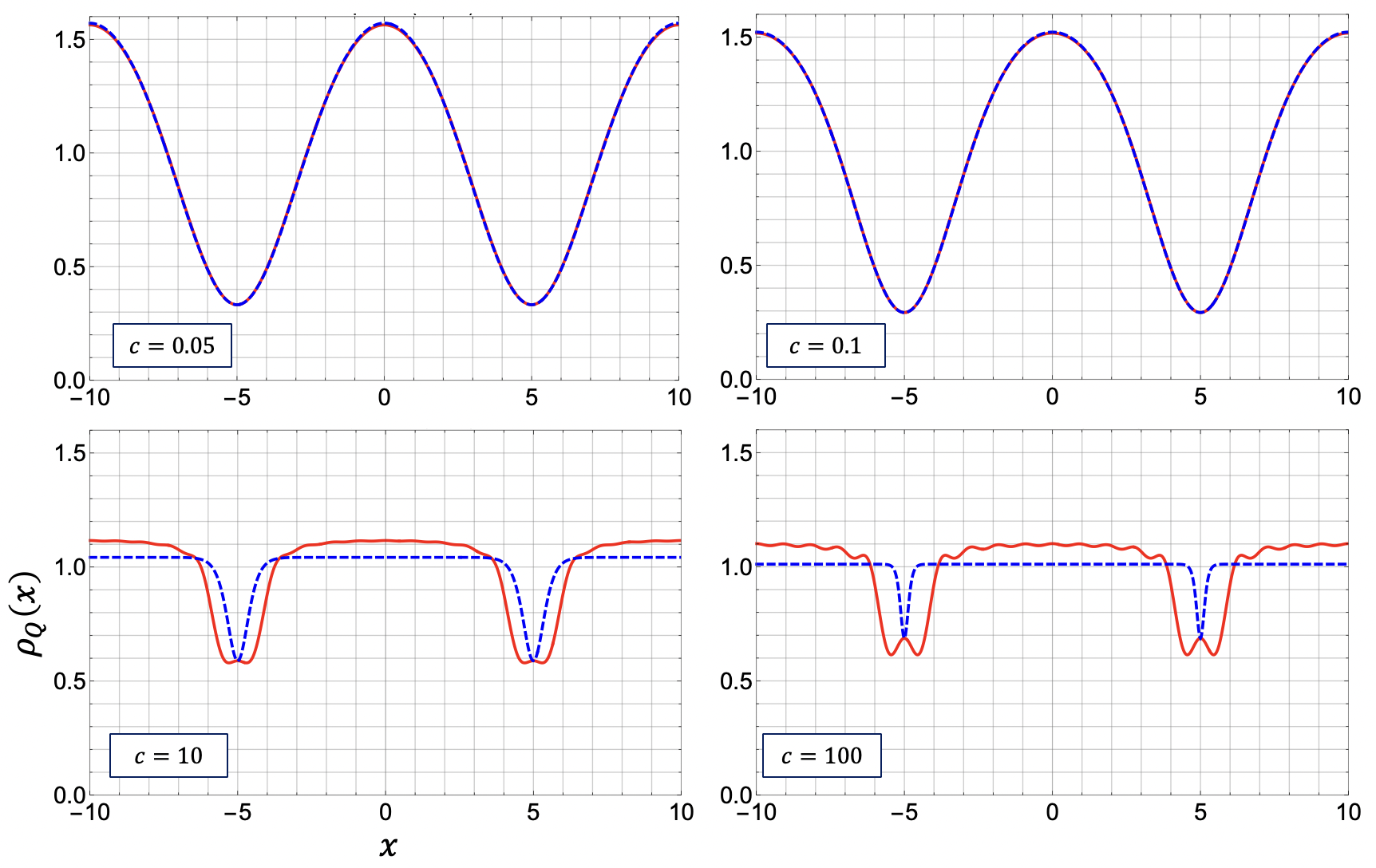}
    \caption{Density profiles of the quantum double dark soliton $\rho_Q(x)$ (red solid curves) and the corresponding elliptic solutions (blue broken curves) for $N=L=20,\ X_1=L/4,\ X_2=-L/4,$ and coupling constants $c = 0.05,\ 0.1,\ 10,\ \text{and } 100$. The parameters of the elliptic solutions are shown in Table \ref{tab:param_bnum_den}.}
    \label{fig:qdoublesol_den}
\end{figure}

It is quite marvelous how completely the density profiles $\rho_Q(x)$ of the quantum double dark solitons in the weak coupling regime agree with those of the corresponding elliptic multiple dark solitons \eqref{eq:rho-EK},  as shown in Figure \ref{fig:qdoublesol_den}.  
It shows the quantum double dark solitons \eqref{eq:2quantum_soliton} and their corresponding elliptic soliton solutions with different coupling constants. The quantum double dark solitons are symmetrical about the line $x=0$ in all cases, reflecting the fact that their quantum states are constructed by a superposition of two-hole excitations with equal weight. 

However, when the coupling constant is not small, each notch of the quantum double dark-soliton is not as deep as the single soliton is. Furthermore, it has two minima for the case $c= 10$ and $100$, as shown in Figure \ref{fig:qdoublesol_den}. An undulating line connects two quantum solitons, which tend to localize more in a larger interaction regime. 

\begin{table}[ht]
\centering
\caption{Parameters of the elliptic solutions for Figure \ref{fig:bnum_den}.}
\begin{tabular}[t]{ccccc}
\hline
$c$ & $k$ & $\mu/2n c$ & $v_c/2\sqrt{nc}$ & $v/v_c$  \\
\hline
0.05 & 0.681 750 & 1.112 62 & 1.657 91 &  0.986 753\\
0.1 & 0.840 623 & 1.115 19 & 1.325 36 & 0.959 493\\
10 & 1- 4.489 79$\times10^{-9}$ & 1.006 60 & 0.998 333 & 0.814 300\\
100 & 1- 1.484 64$\times10^{-24}$ & 1.001 34 & 0.999 156 & 0.841 665\\
\hline
\end{tabular}
\label{tab:param_bnum_den}
\end{table}%

\subsubsection{Matrix elements of the quantum double dark soliton state}
We have calculated the matrix element by using the determinant formula \eqref{eq:Slavnov_field}. The matrix element $\psi_Q(x)$ is given by
\begin{align}
    \begin{split}
    &\psi_Q(x):=\braket{X_1,X_2,N-1|\hat{\psi}(x)|X_1,X_2,N}\\
	=&\frac{1}{\mathcal{M}}\sum_{\bm{p},\bm{p'}\in \bm{P}} e^{i(P-P')x} e^{i(p_1X_1+p_2X_2)}e^ {-i(p'_1X_1+p'_2X_2)}\braket{P_1,P_2,N-1|\hat{\psi}(0,0)|P_1',P_2',N},
	\label{eq:2quantum_soliton_matele}
\end{split}
\end{align}
 which takes complex value for position $x$. The classical counterpart of the matrix element $\psi_Q(x)$ is the relevant elliptic solution which is given by Eq. \eqref{eq:ansatz_esol}, with its absolute part and argument. Comparisons of the matrix element $\psi_Q(x)$ and the classical soliton $\psi_C(x)$ are made by plotting the square amplitudes and the arguments.
 Note that the density $n$ is not always $N/L$, since the number of particles in the matrix element is not conserved.
 
\begin{figure}[H]
    \centering
    \includegraphics[width=14cm]{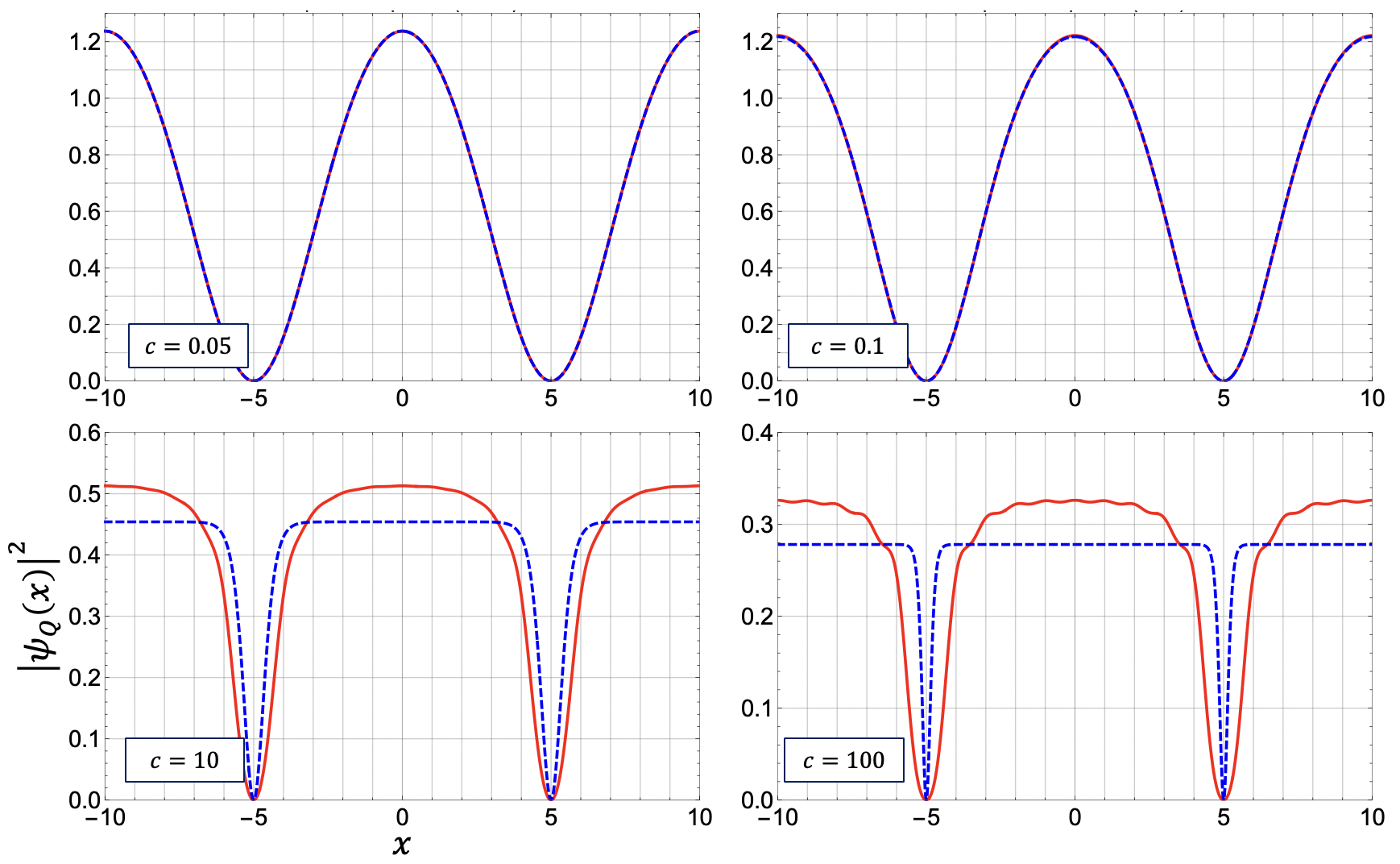}
    \caption{Square amplitudes $|\psi_Q(x)|^2$ for the matrix elements \eqref{eq:2quantum_soliton_matele} (red solid curves) and their classical counterparts (blue broken curves) for coupling constants $c=0.05,\ 0.1,\ 10,$ and 100. The parameters of the elliptic solutions are shown in Table \ref{tab:param_matele}.}
    \label{fig:2qs_sq_amp}
\end{figure}

 Figure \ref{fig:2qs_sq_amp} shows the profile of the square amplitude $|\psi_Q(x)|^2$ for the matrix element \eqref{eq:2quantum_soliton_matele}. The smallest values of the square amplitude of the notches are smaller than those of the density profiles in Figure \ref{fig:bnum_den}. In the case of $c = 0.05$ and 0.1, the square amplitudes shown in Figure \ref{fig:2qs_sq_amp} almost completely agree with the corresponding elliptic solutions. In the case of $c = 10$ and 100, however, the widths of the dips still remain wider than those of the elliptic solutions.
 
\begin{figure}[H]
    \centering
    \includegraphics[width=13cm]{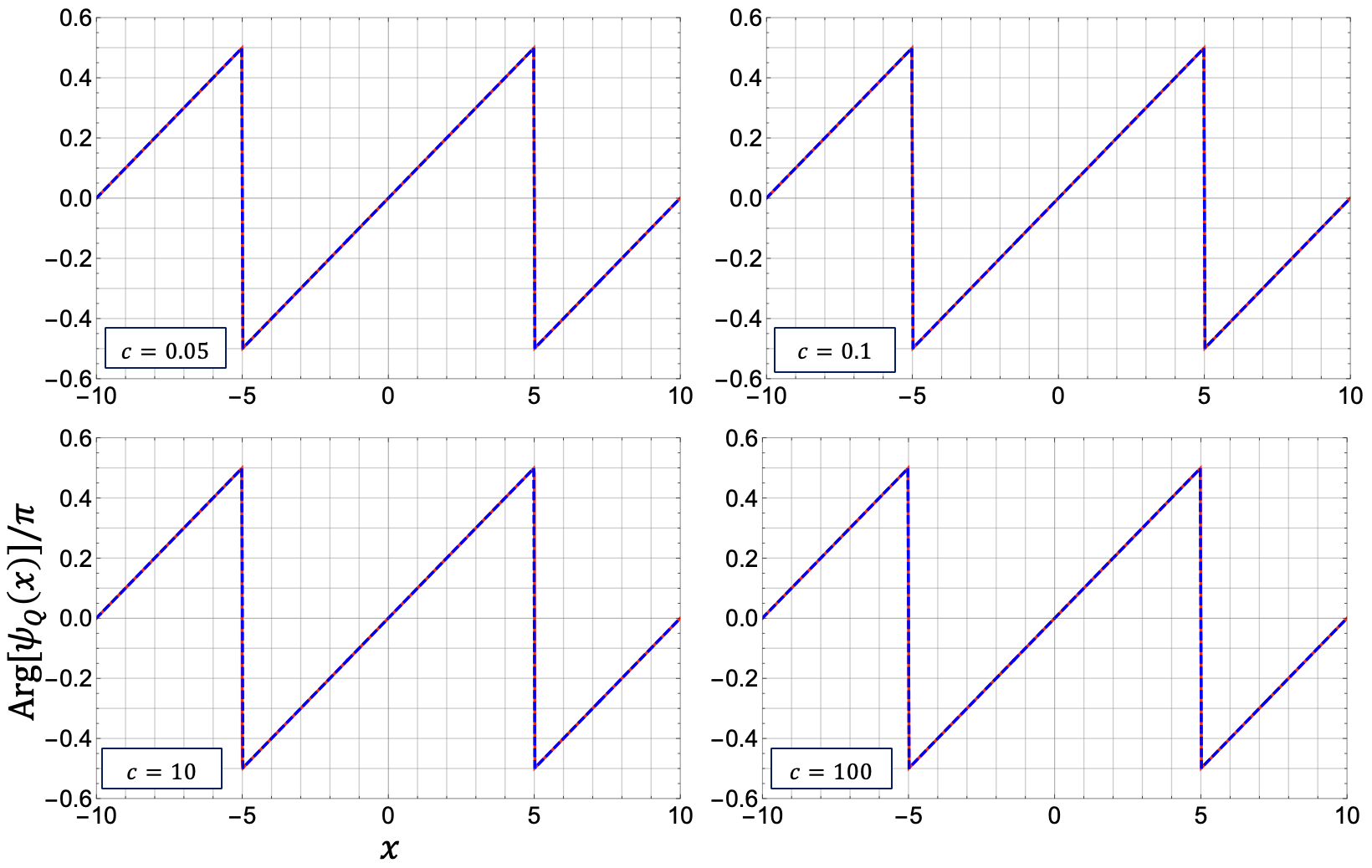}
    \caption{Phase profiles $\arg[ \psi_Q(x) ]/\pi$ for the matrix elements \eqref{eq:2quantum_soliton_matele} (red solid curves) and their classical counterparts (blue broken curves) for coupling constants $c=0.05,\ 0.1,\ 10,$ and 100. The parameters of the elliptic solutions are shown in Table \ref{tab:param_matele}. In the all panels, the red solid curves completely agree with the blue broken curves.}
    \label{fig:2qs_arg}
\end{figure}

 Figure \ref{fig:2qs_arg} shows the phase profile $\arg[\psi_Q(x)]$ for the matrix element \eqref{eq:2quantum_soliton_matele}.
The phase profiles in Figure \ref{fig:2qs_arg} agree with those of the elliptic solutions for all coupling constants. It is noted that the phase profiles with four different coupling constants $c$ shown in Figure \ref{fig:2qs_arg} are similar for both quantum and classical cases. We can confirm that the phase profile $\arg[\psi_Q(x)]$ is satisfied with the periodic boundary condition such as $\arg[\psi_Q(x+L)]= \arg[\psi_Q(x)]$ for the period $L$. The abrupt jumps at the solitons positions in Figure \ref{fig:2qs_arg} are consistent with our previous result which says that the deep notch in a soliton amplitude gives the abrupt jump in its phase profile \cite{Sato_2016}.

The phase profile of the notch located at $x=5$ is completely the same as that of $x=-5$. The phase profile in total is point symmetric with respect to the origin $x=0$.


\begin{table}[ht]
\centering
\caption{Parameters of the elliptic solutions for Figures \ref{fig:2qs_sq_amp} and \ref{fig:2qs_arg}.}
\begin{tabular}[t]{cccccc}
\hline
$c$ & $k$ &  $n$  &$\mu/2n c$ & $v_c/2\sqrt{nc}$ & $v/v_c$  \\
\hline
0.05 & 0.681 520  &  0.666918  & 1.44 586 & 1.75 158 &  0.982 197\\
0.1  & 0.838 694  &  0.699671  & 1.40214 & 1.267 41 & 0.937 102\\
10   & 1- 4.444 99 $\times10^{-9}$ &  0.411534  & 1.091 57 & 0.916 966 & 0.168 885\\
100  & 1- 9.964 63$\times10^{-23}$ &   0.267601 &  1.037 57 & 0.963 609 & 0.0630 238\\
\hline
\end{tabular}
\label{tab:param_matele}
\end{table}%

\subsection{Quantum states of a double dark soliton with Gaussian weights}\label{subsec_2qs_g}

\subsubsection{Gaussian weights in terms of target soliton depth}

We now demonstrate that by assigning the Gaussian weights 
to the sum of eigenstates in Eq. \eqref{eq:2quantum_soliton_state} 
quantum states for double dark-solitons are quite effectively 
constructed. The method is similar to that of constructing quantum single dark soliton states \cite{PhysRevA.99.043632(Shamailov-Brand2019),kaminishi2020construction}: We assign the ideal Gaussian weights 
\cite{kaminishi2020construction} to the sum of the eigenstates 
with two particle-hole excitations given in Eq. \eqref{eq:2quantum_soliton_state}. Then, the constructed quantum states give such density profiles that have distinct notches with different depths. 

We employ the ideal Gaussian weights \cite{kaminishi2020construction} 
explained in section  \ref{subsec_singledarksoliton_g}.  
Hereafter in this subsection we express them in terms of the target soliton depth denoted by $d$.  We shall define it shortly. 
%
By assigning a target soliton depth to each of the two notches, we construct a quantum double soliton state such that its density profile has two distinct notches with different depths.  

Let us consider the Gaussian superposition of the excited states consisting of two particle-hole excitations which are determined by a pair of holes $\bm{p} = \{p_1, p_2\}$ in the set $\bm{P}$:
\begin{align}
    \ket{X_{1}, X_{2}, N}_G =\frac{1}{\sqrt{\mathcal{N}}} \sum_{\{p_1, p_2\}\in \bm{P}} G_{P_0}(p_1) G_{P'_0}(p_2) e^{i (p_{1} X_{1}+ p_{2} X_{2})}\left|p_{1}, p_{2}, N\right\rangle.
    \label{eq:2quantum_soliton_state_g}
\end{align}
Here, $\mathcal{N}$ is a normalization factor and 
the set $\bm{P}$ is the same as given in section \ref{subsec_qdoublestate}. The Gaussian function is given by
\begin{align}
    G_{P}(q)= \exp \left[-\frac{\left(q-P\right)^{2}}{4 \sigma^2}\right]
\end{align}
with two Gaussian parameters $(P,\sigma)$. 
They are determined by target soliton depth $d$, as we shall show shortly. 

We now define the target soliton depth $d$. 
It is related to the ratio ${v}/{v_{c,\infty}}$ 
through the dark soliton solution in the thermodynamic limit \cite{kaminishi2020construction}
and defined by the following: 
%
\begin{align}
\sqrt{d} =|\phi_P^\infty(0)|=\sqrt{n}\left|\frac{v}{v_{c , \infty}} \right|.
\label{eq: target_depth}
\end{align}
Here $|\phi_P^\infty(0)|$ denotes the square root 
of the local density $\rho(x)$ at the origin, i.e., at the position of a notch, in the thermodynamic limit. We recall that the local density $\rho(x)$ near a notch has been given in Eq. \eqref{eq:j-th-amp}. We also recall that the critical velocity $v_{c , \infty}$ has been defined in section  \ref{subsub:reduction}. 

In terms of the target soliton depth $d$ and density $n=N/L$, the Gaussian parameters $P$ and $\sigma$ are given by 
\begin{align}
    P(d) &= 2 n \left\{ \frac{\pi}{2}-\left[ \sqrt{\frac{d}{n}\left(1-\frac{d}{n}\right)} +\arcsin{\left(\sqrt{\frac{d}{n}}\right)} \right] \right\} \, ,  \label{eq:g_param_p}
    \\
    \sigma^2 & =\frac{4}{3} n \sqrt{n c}\left( 1-\frac{d}{n}\right) ^\frac{3}{2} \label{eq:g_param_sigsq}.
\end{align}

\subsubsection{Profiles of density \texorpdfstring{$\rho_Q(x)$}{} and square amplitude \texorpdfstring{$|\psi_Q(x)|^2$}{}
}

\begin{figure}[H]
    \centering
    \includegraphics[width=13cm]{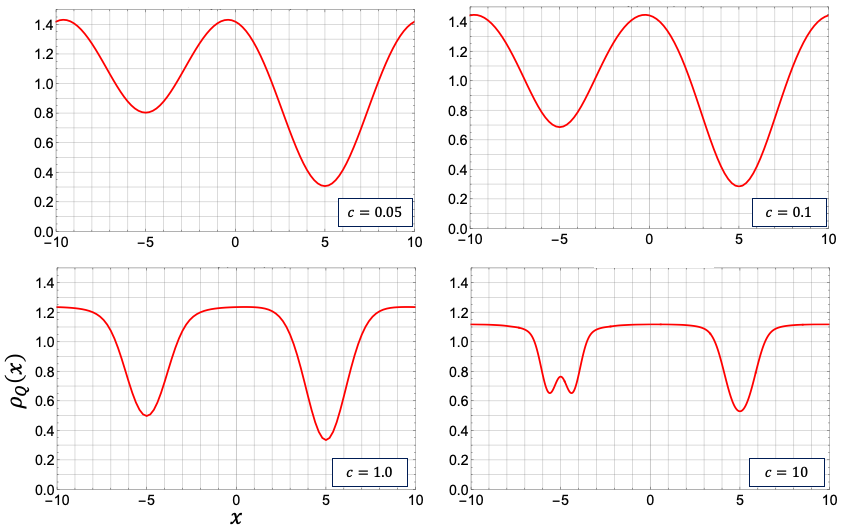}
    \caption{Density profiles $\rho_Q(x)$ for $N=L=20$ and $c=0.05,\ 0.1,\ 1.0,\ \text{and}\ 10$. The target soliton depths $d$ for the Gaussian weights are given by $0.6$ and $0$, and the Gaussian parameters are given by $(P_0, \sigma) =  (0.124027\pi ,0.106667)$ and $(P'_0, \sigma') = (\pi ,0.421637)$, which are calculated by Eqs \eqref{eq:g_param_p} and \eqref{eq:g_param_sigsq}. The density profile is calculated in the same way as in Eq. \eqref{eq:2quantum_soliton}.}
    \label{fig:2qs_den_g}
    \centering
\end{figure}

\begin{figure}[H]
    \centering
    \includegraphics[width=13cm]{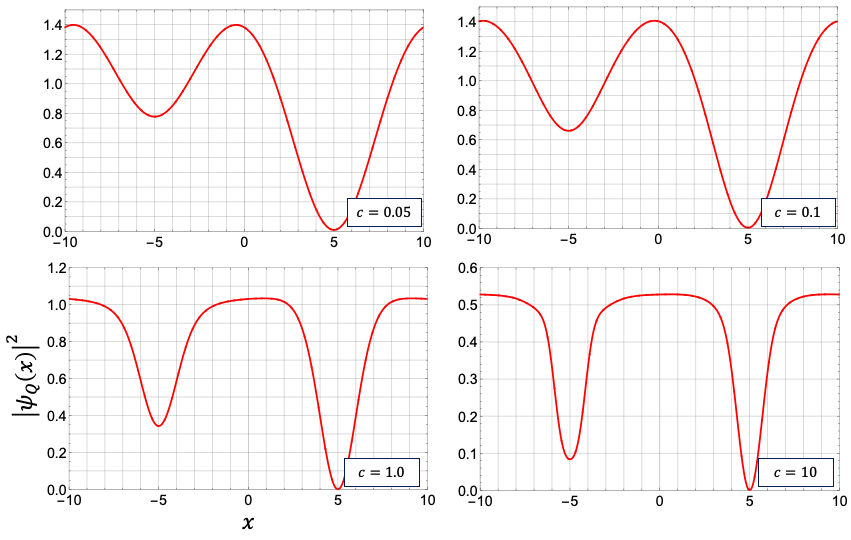}
    \caption{Square amplitudes of the matrix element, $|\psi_Q(x)|^2$, for $N=L=20,$ and $c=0.05,\ 0.1,\ 10,\ \text{and}\ 100$. The pairs of the Gaussian parameters are given by the same sets of values as in Figure \ref{fig:2qs_den_g}.}
    \label{fig:2qs_sq_amp_g}
\end{figure}

For the quantum state of Eq. \eqref{eq:2quantum_soliton_state_g} 
we exactly compute the density $\rho_Q(x)$ and the matrix element $\psi_Q(x)$ and plot the profiles of the density $\rho_Q(x)$, the square amplitude $|\psi_Q(x)|^2$ and the phase $\arg[\psi_Q(x)]$ in Figures \ref{fig:2qs_den_g},\ref{fig:2qs_sq_amp_g},
and \ref{fig:2qs_arg_g}, respectively, 
for several values of the coupling constants.
%
%
For all the figures in this subsection, 
the system length $L$ and the number of particles $N$ are 
given by $N=L=20$. The target soliton depths are set as $d=0.6$ and $d'=0$ for each notch, so that the pairs of Gaussian parameters in the quantum state of Eq. \eqref{eq:2quantum_soliton_state_g} are given by $(P_0, \sigma) =  (0.124027\pi ,0.106667)$ and $(P'_0, \sigma') = (\pi ,0.421637)$.

In the density profiles of Figure \ref{fig:2qs_den_g} we observe double dark-solitons consisting of two distinct narrow notches with different depths in the two panels of the weak coupling regime for $c \le 0.1$. 
In the two panels of $c \le 0.1$ it is clear that the depths of the two notches are different. 
We consider it a consequence of assigning the Gaussian weights to the sum of  eigenstates with two particle-hole excitations. 
%
Thus, the plots in Figure \ref{fig:2qs_den_g} confirm that the construction with the ideal Gaussian weights reviewed in section \ref{subsec_singledarksoliton_g} is effective in the weak interaction regime such as $c\leq 0.1$.

%



We thus suggest from Figures \ref{fig:2qs_den_g} and  \ref{fig:2qs_sq_amp_g}
that the Gaussian superposition of the eigenstates with two particle-hole excitations in Eq. \eqref{eq:2quantum_soliton_state_g} 
should produce a quantum state rather close to the mean-field product state for a given double dark-soliton, when the coupling constant $c$ is small enough such as $c \le 0.1$.  
In particular, the almost complete agreement between the density profiles in Figure \ref{fig:2qs_den_g} and the profiles of the square amplitude in Figure \ref{fig:2qs_sq_amp_g} for $c \le 0.1$ suggests that the quantum state with the ideal Gaussian weights in Eq. \eqref{eq:2quantum_soliton_state_g} should be close to the corresponding mean-field product state. 
In the panels of $c=0.05$ and $0.1$, every notch in Figures \ref{fig:2qs_den_g} and \ref{fig:2qs_sq_amp_g} is distinct and narrow. 
Furthermore, the value of $\rho_Q(x)$ or $|\psi_Q(x)|^2$ at the starting points of the dips, i.e., the maximal points in the profiles,  one of which is located around at $x=0$ for instance, is given by 1.4 in both Figures \ref{fig:2qs_den_g} and \ref{fig:2qs_sq_amp_g}. 

Here we remark that if the Gaussian superposition of the eigenstates in the Lieb-Liniger model of Eq. \eqref{eq:2quantum_soliton_state_g} completely 
produce the mean-field product state satisfying Eq. \eqref{eq:soliton_wave}, 
then the square amplitude of the matrix element $\psi_Q(x)={}_G\braket{X_1, X_2,N-1|\hat{\psi}(x)|X_1, X_2,N}_G$ is equal to the density profile $\rho_{Q}(x) ={}_G\braket{X_1, X_2,N|\hat{\psi}^\dagger(x)\hat{\psi}(x)|X_1, X_2,N}_G$: 
We have $\rho_Q(x) = |\psi_Q(x)|^2$. 
%


\subsubsection{Difference between density \texorpdfstring{$\rho_Q(x)$}{} 
and square amplitude \texorpdfstring{$|\psi_Q(x)|^2$}{}}
\label{sec:difference}

We now estimate how far the quantum soliton state constructed by Eq.  \eqref{eq:2quantum_soliton_state_g} satisfies the characteristic property of Eq.  \eqref{eq:soliton_wave} for the  mean-field product state. 
Let us subtract the square amplitude of the matrix element, $|\psi_Q(x)|^2$, 
from the density, $\rho_Q(x)$.
Inserting the identity operator between $\hat{\psi}^\dagger(x)$ and $\hat{\psi}(x)$ in the density operator, we obtain the square amplitude of the matrix element and other terms:
\begin{align}
\begin{split}
    &{}_G\braket{X_1, X_2,N|\hat{\psi}^\dagger(x)\hat{\psi}(x)|X_1, X_2,N}_G\\
     =&\sum_\chi {}_G\braket{X_1, X_2,N|\hat{\psi}^\dagger(x)\ket{\chi}\bra{\chi}\hat{\psi}(x)|X_1, X_2,N}_G\\
     =& |{}_G\braket{X_1, X_2,N-1|\hat{\psi}(x)|X_1, X_2,N}_G|^2
    +|\braket{\chi'|\hat{\psi}(x)|X_1, X_2,N}_G|^2+\cdots.
    \label{eq:den_sa_relation}
\end{split}
\end{align}
We define the difference $D(x)$ by 
 \begin{align}
 D(x) :=
     &{}_G\braket{X_1, X_2,N|\hat{\psi}^\dagger(x)\hat{\psi}(x)|X_1, X_2,N}_G - |{}_G\braket{X_1, X_2,N-1|\hat{\psi}(x)|X_1, X_2,N}_G|^2.
 \end{align}
Since each term on the right hand side of Eq. \eqref{eq:den_sa_relation} is non-negative,  we have the relation 
\begin{equation}
    D(x) \geq 0 . 
\end{equation}

The difference $D(x)$ expresses how far the quantum soliton state is  different from the corresponding mean-field product state which should satisfy the property of Eq. \eqref{eq:soliton_wave}. Moreover, we remark that $D(x)$ depends on the position $x$ and the coupling constant $c$. 

We now argue that the quantum double dark soliton state of Eq.  \eqref{eq:2quantum_soliton_state_g} should be close to the 
corresponding mean-field product state in the weak coupling regime such as $c \le 0.1$. 
Let us evaluate $D(x)$ for weak and strong coupling constants.
When the coupling constant $c=0.05$, the difference $D(x=0)$ is almost zero, while $D(x=5)=0.3$, from Figure  \ref{fig:2qs_den_g} and \ref{fig:2qs_sq_amp_g} (see also Table \ref{tab:difference}). 
On the other hand, for a strong coupling constant $c=10$, the difference $D(x)$ is about 0.6 except for the notches.

Here we recall that in both Figures \ref{fig:2qs_den_g} and \ref{fig:2qs_sq_amp_g}  around at $x=0$ one of the two notches starts to dip, i.e.,  the value of $\rho_Q(x)$ or $|\psi_Q(x)|^2$ decreases, 
and it is approximately given by 1.4, as shown in Table \ref{tab:difference}.

\begin{table}[h]
 \caption{The values of the density profile$\rho_Q(x)$, those of the square amplitudes $|\psi_Q(x)|^2$, and their differences $D(x)$ at $x=0$ and $x=5$. }
 \label{table:ex_booktabs}
 \centering
  \begin{tabular}{lrrrcrrr}
   \toprule             
   $c$  & \multicolumn{3}{c}{$x=0$}& \phantom{a} & \multicolumn{3}{c}{$x=5$}\\
   \cmidrule(lr){2-4} \cmidrule{6-8} 
       & $\rho_Q(x)$ & $|\psi_Q(x)|^2$ & $D(x)$ & & $\rho_Q(x)$ & $|\psi_Q(x)|^2$ & $D(x)$  \\
   \midrule    
   0.05 & 1.41865 & 1.38161 & 0.0370444 & & 0.306279 & $9.80563\times10^{-3}$ & 0.296473\\
   0.1 & 1.44277 & 1.40292 & 0.0398479 & & 0.283943 & $4.46456\times10^{-3}$ & 0.279478\\
   1.0 & 1.23462 & 1.03089 & 0.203732 & & 0.334114 & $4.24589\times10^{-6}$ & 0.334109\\
   10 & 1.11822 & 0.528072 & 0.59015 & & 0.528875 & $1.96588\times10^{-4}$ & 0.528678 \\
   \bottomrule          
  \end{tabular}
\label{tab:difference}
\end{table}

\subsubsection{Phase profile of the quantum double soliton state}

\begin{figure}[H]
    \centering
    \includegraphics[width=13cm]{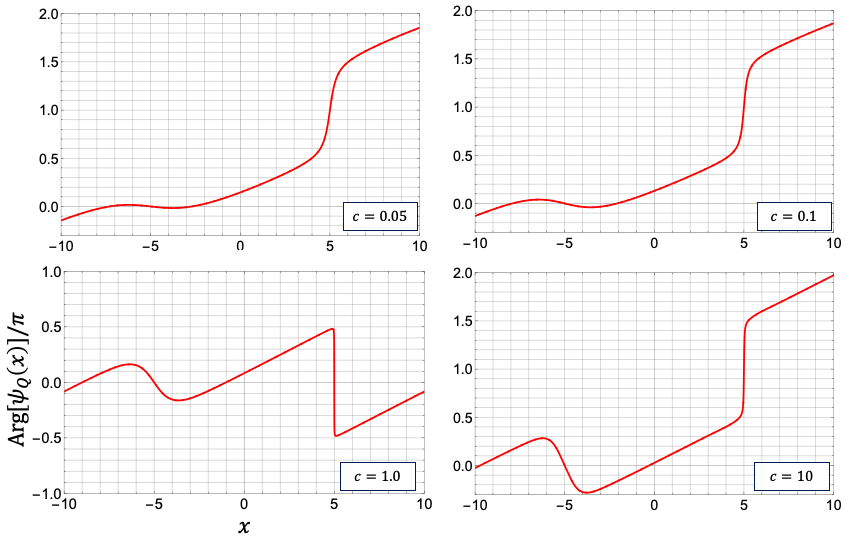}
    \caption{Phase profiles $\arg[\psi_g(x)]/\pi$ for $N=L=20,$ and $c=0.05,\ 0.1,\ 1.0,\ \text{and}\ 10$. The pairs of the parameters of the Gaussian function are the same as Figure \ref{fig:2qs_den_g}.}
    \label{fig:2qs_arg_g}
\end{figure}

The phase profiles in Figure \ref{fig:2qs_arg_g} are different from those of no Gaussian weights plotted in Figure \ref{fig:2qs_arg}
with respect to the point symmetry and the boundary conditions.   
The phase profiles do not have the point symmetry at the origin 
in all the  panels of  the different values of the coupling constant in Figure \ref{fig:2qs_arg_g}. Furthermore, except for the panel of $c=1.0$, the phase profile satisfies the boundary condition: $\arg[\psi_g(x+L)] = \arg[\psi_g(x)] + 2 \pi$. 
The phase profile is shifted by $2\pi$ at the ends, 
while it is continuous with respect to the coordinate $x$.

Assigning the Gaussian weights leads to  the change of the winding numbers for the corresponding solitons. 
For an illustration, let us consider the upper-right panel 
for $c=0.1$ of Figure \ref{fig:2qs_arg_g}.  
We can show that the winding number of one soliton with its notch located at $x=-5$ is given by $J=0$, 
while that of another located at $x=5$ is given by $J=1$. 
We shall explain it  later in detail. 
%

\begin{figure}[H]
    \centering
    \includegraphics[width=8cm]{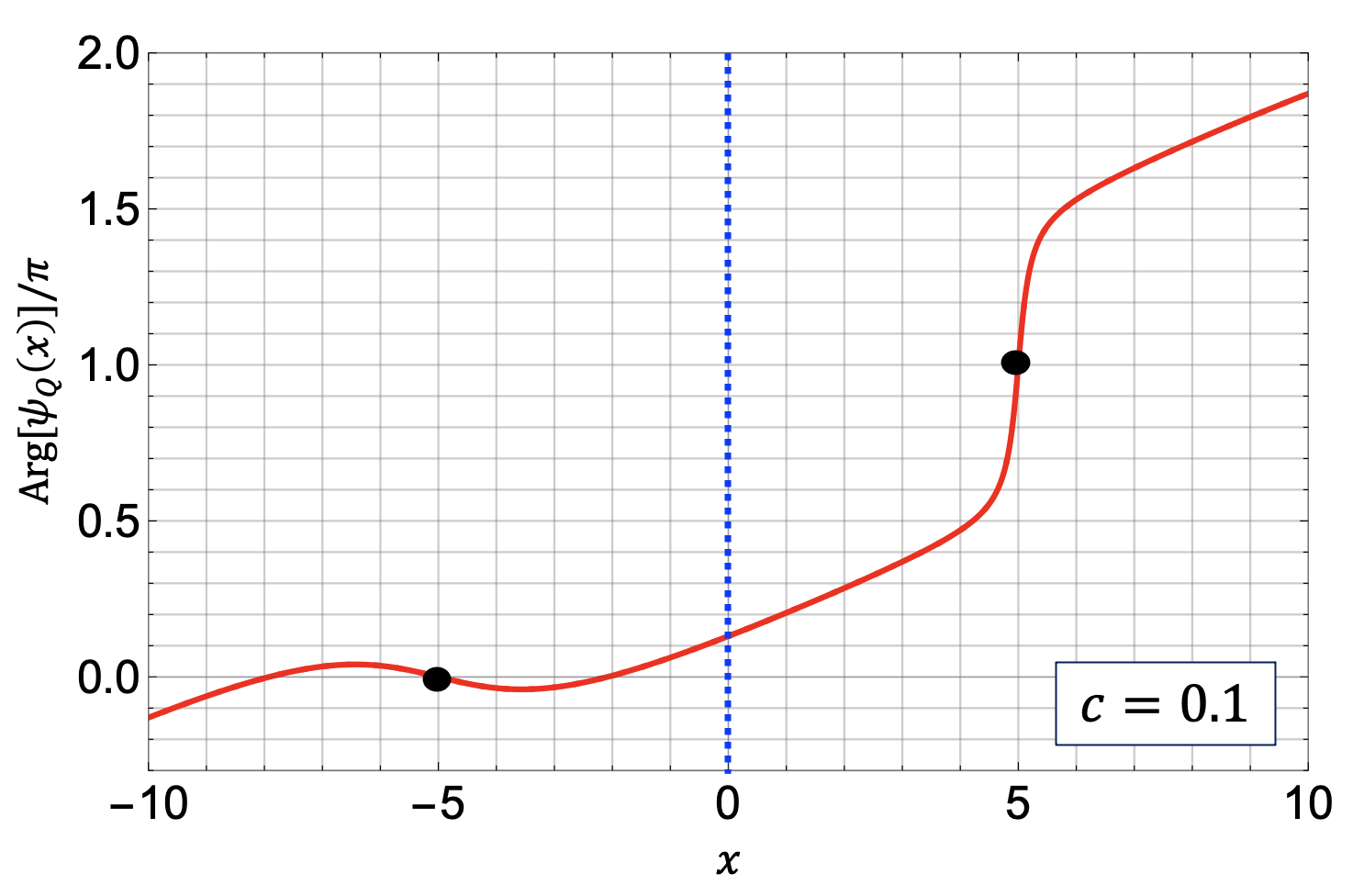}
    \caption{Phase profile $\arg[\psi_g(x)]/\pi$ (red curve) for $N=L=20$ and $c=0.1$ divided by the vertical line  $x=0$ (blue line).  }
    \label{fig:PhaseProfile_c0.1}
\end{figure}

We decompose the phase profile in the upper-right panel for $c=0.1$ of Figure \ref{fig:2qs_arg_g} around at $x=0$ into two phase sub-profiles of single dark solitons: one sub-profile in the region of $x<0$ with an N-shaped figure centered at $x=-5$, and another one in the region of $x>0$ with an abrupt jump at $x=5$. In Figure \ref{fig:PhaseProfile_c0.1} the phase profile is divided by the vertical line at $x=0$ to two sub-profiles in the regions of $x<0$ and $x>0$, respectively. The black dots depict the centers of the phase sub-profiles. Here, they are located at $x=-5$ and 5, respectively.

We shall argue that the winding numbers of the two sub-profiles are given by $J=0$ and $J=1$, respectively. 
Here we recall that the phase field is approximately expressed by Eq. \eqref{eq:abrupt} if the system size $L$ is very large. We also recall that elliptic multiple  dark-soliton solutions are described in section \ref{subsec_ellipticmulti}. The phase field $\varphi(x)$ is expressed in terms of the elliptic 
integral of the third kind together with velocity $v$ and  parameter $W$ in Eq. \eqref{eq:phase} or more compactly 
in Eq. \eqref{eq:phase-theta} in terms of the elliptic theta function. The parameter $v$ is determined by the boundary condition \eqref{eq:phase_bc} with the winding number $J$. For the real parameter $W$, we have the choice of signs, as shown in Eq. \eqref{eq:W}.

\begin{figure}[H]
    \centering
    \includegraphics[width=15cm]{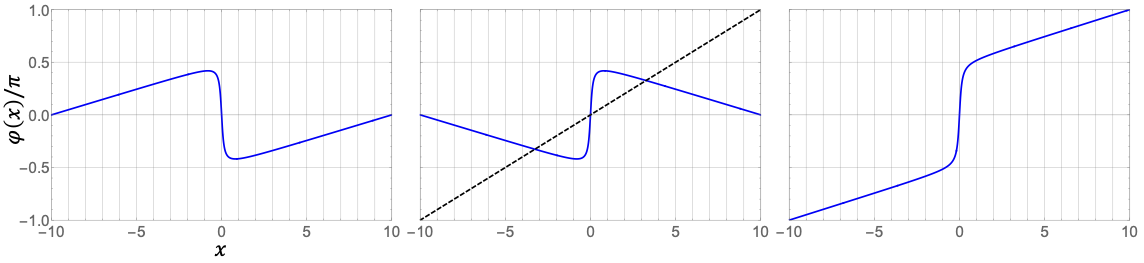}
    \caption{Phase profiles of the elliptic soliton solution of Eq. \eqref{eq:phase}. For $J=0$, $W<0$ and hence $v >0$ (in the leftmost panel); for $J=0$, $W>0$ and hence $v<0$ (in the middle panel); for $J=1$ and $W>0$ (in the rightmost panel). The black broken line $\frac{2 \pi x}{L}$ illustrates the first term of the phase profile in Eq. \eqref{eq:phase-theta} which is also contained in Eq. 
    \eqref{eq:phase}. }
    \label{fig:PhaseProfile_explain}
\end{figure}

The left-hand side ($x < 0$) of the phase profile in the upper-right panel ($c=0.1$) of Figure \ref{fig:2qs_arg_g} has the same shape as the N shaped profile in the leftmost panel of Figure \ref{fig:PhaseProfile_explain}. The unwinding phase profile ($J=0$) for $W<0$ and that of $W>0$ are exhibited in the two panels from the left in Figure \ref{fig:PhaseProfile_explain}, respectively,  expressed as blue curves. Thus, we suggest that the left-hand side phase profile in the region of $x< 0$ in the upper-right panel ($c=0.1$) of Figure \ref{fig:2qs_arg_g} should correspond to the phase profile of a single elliptic dark-soliton with $J=0$ and $W<0$. Here we recall that the abrupt decrease ($W<0$) (or increase ($W>0$)) near the center ($x=0$) of the profile is explained in section \ref{sec:abrupt_change}. 

On the other hand, the right-hand side phase profile ($x > 0$) in the upper-right panel ($c=0.1$) of Figure \ref{fig:2qs_arg_g} should correspond to the phase profile of a single elliptic dark-soliton with $J=1$ and $W>0$. When the winding number $J=1$, the term $4\pi/L$ is added to the velocity for $J=0$ due to the boundary condition. It follows that the $J=1$ phase profile is given by the superposition of the linear graph of $2\pi x/L$ and the $J=0$ phase profile. In the middle panel of Figure \ref{fig:PhaseProfile_explain}, the line graph $\frac{2 \pi}{L} x$ (black broken line) and the phase profile with $J=0$ and $W>0$ (blue curve) are plotted. 
The resultant winding phase profile with $J=1$ is shown 
in the rightmost panel of Figure \ref{fig:PhaseProfile_explain}. 
We suggest that it should correspond to 
the right-hand side phase profile ($x > 0$) 
in the upper-right panel of Figure \ref{fig:2qs_arg_g}.

Thus, we have observed that by assigning the Gaussian weights with certain parameters we obtain not only the density profiles consisting of two notches with different soliton depths but also the phase profiles consisting of two dark-solitons with different winding numbers.


\section{Conclusion}
We have introduced novel quantum states of single dark solitons with a nonzero winding number. Then, we have constructed the quantum double dark-soliton states by taking the analogy to the case of the free fermions and the construction of quantum single dark soliton states with and without Gaussian weights.
The superposition of the eigenstates with 
two particle-hole excitations in the Lieb-Liniger model leads to such quantum states that have two notches in the density profile as a double dark-soliton.
Two notches with the same depths appear in the superposition of two particle-hole excitations with equal weight. Two notches with different depths are generated by those of non-equal weights, in particular, the ideal Gaussian weights. 


For quantum double dark solitons, we have controlled their depths by some parameters. It has never been performed in the previous studies, as far as we have investigated. We have obtained the quantum double dark soliton which possesses distinct and narrow notches with controlled depths in both density profiles and the profiles of the square amplitude of the matrix element of the field operator, in the case when $N=L=20$ and $c=0.05$ and $0.1$.

Furthermore, we have confirmed that the quantum double dark soliton state satisfies the assumption of the mean-field product state for small coupling constant by evaluating the difference $D(x)$. That is, the density profiles are close to the square amplitudes of the matrix elements of the field operator if the coupling constant is small enough, such as $c<0.1$ for $N=20$.



As an important topic we have shown that the ansatz of Ref. ~\cite{kaminishi2020construction} is effective and useful as far as the density profile is concerned. The ansatz was also quite useful for constructing double dark soliton states as we have shown in \S 4. 
However, we should point out that it does not completely lead to the mean-field product state. In fact, one of the key properties of the mean-field product state, i.e., Eq. (\ref{eq:soliton_wave}), 
holds only approximately for all the examples investigated in this article. 
Furthermore, the phase field of the constructed quantum state sometimes has a nonzero winding number, i.e., it satisfies one of non-periodic boundary conditions.  
It is maybe due to the finite-size effect. We expect that one can improve the construction so that one obtain a complete mean-field product state by increasing the number of particles $N$ and the system size $L$, while by decreasing the coupling constant $c$.

\section{Discussion}

There is another approach to observe quantum dark-solitons numerically: successive measurements of particle positions in the Lieb-Liniger model studied by Syrwid et. al \cite{PhysRevA.92.032110(Syrwid-Sacha2015), Syrwid2016}. They showed quantum single/double dark-solitons by measuring particle positions in a single one-/two-hole excitation state. Even though their method are different from ours in section \ref{singledarksoliton} and \ref{subsec_qdoublestate}, both studies share the results that the phase profiles have the abrupt jumps at the solitons positions, and the two notches in a double soliton have the same depths.

The question is how the density profile of a superposition of yrast states are associated with the successive measurements of particle positions.
For the quantum single dark-soliton case, some groups worked on this question. When a coupling constant $c = 0$, it was analytically shown the relationship between the construction with Gaussian weight \cite{kaminishi2020construction} and the particle position method \cite{PhysRevA.92.032110(Syrwid-Sacha2015)} in Ref. \cite{kaminishi2020construction}. In the case of a small coupling constant $c > 0$, an ansatz was proposed to bridge between the calculation of single-particle density and the particle position method, in a recent research \cite{Golletz2020}. 

On the other hand, for the quantum multi dark-soliton case, the connection between the calculation of single-particle density and the particle position method is still unclear. We suggest that mean-field state with the double dark soliton solution of the NS equation could reveal the connection. The connection is still unclear, but we showed some quantum states associated with multiple dark-solitons and their properties explicitly.

We leave the comparison of the quantum double dark-soliton with Gaussian weights and its classical counterpart for future work. The system we consider here is under the periodic boundary conditions, while the soliton solution of the NS eq., obtained by the inverse scattering method does not satisfy the periodic boundary conditions. The finite-gap soliton \cite{belokolos1994algebro} should be the one to compare with the quantum double dark-soliton with Gaussian weights. For the appropriate parameters, the finite-gap soliton owns two notches with the different depths. We will find the finite-gap solitons which correspond to the quantum double dark-soliton with Gaussian weights.

Moreover, we shall study the dynamics of the quantum dark solitons, applying the construction with Gaussian weights to them. We have not yet observed the dynamics of the quantum double dark soliton even if both of the solitons are black. The dynamics of the quantum dark solitons should reveal soliton properties such as phase shifts.

\vspace{6pt} 




\funding{The present research was partially supported by the Japan Society for the Promotion of Science (KAKENHI Grants-in-Aid for Scientific Research: Grant No. 21K03398). E.K. was supported by JSPS Grant Number 20K14388. }

 \acknowledgments{
}

\conflictsofinterest{The authors declare no conflict of interest.} 





\reftitle{References}

\externalbibliography{yes}
\bibliography{bibtex}




\end{document}